\documentclass[AMS,STIX1COL]{WileyNJD-v2}
\usepackage{graphicx}
\usepackage{moreverb}
\usepackage{amsmath}
\usepackage{amssymb}
\usepackage{caption}
\usepackage{subcaption}
\usepackage{mathtools}
\usepackage{epsfig}
\usepackage{pifont}
\usepackage{wrapfig}
\usepackage{xcolor}
\newcommand{\net}{\mathcal{G}}
\newcommand\BibTeX{{\rmfamily B\kern-.05em \textsc{i\kern-.025em b}\kern-.08em
T\kern-.1667em\lower.7ex\hbox{E}\kern-.125emX}}
\articletype{Review article}

\received{<day> <Month>, <year>}
\revised{<day> <Month>, <year>}
\accepted{<day> <Month>, <year>}

\begin{document}

\title{Robustness and Exploration of Variational and Machine Learning Approaches to Inverse Problems: An Overview}

\author[1]{Alexander Auras*$^\dagger$}
\author[1]{Kanchana Vaishnavi Gandikota$^\dagger$}
\author[2]{Hannah Droege}
\author{Michael Moeller$^1$\\$^\dagger$ \addressfont These authors contributed equally}

\authormark{Auras \textsc{et al}}

\address[1]{\orgdiv{Institute for Vision and Graphics}, \orgname{University of Siegen}, \orgaddress{\state{NRW}, \country{Germany}}}

\address[2]{\orgdiv{Institute of Computer Science}, \orgname{Rheinische Friedrich-Wilhelms-Universität Bonn}, \orgaddress{\state{NRW}, \country{Germany}}}

\corres{*Alexander Auras,\\Institute for Vision and Graphics,\\University of Siegen,\\Adolf-Reichwein-Straße 2a,\\57076 Siegen,\\
Germany.\\\email{alexander.auras@uni-siegen.de}}

\abstract[Abstract]{This paper provides an overview of current approaches for solving inverse problems in imaging using variational methods and machine learning. A special focus lies on point estimators and their robustness against adversarial perturbations. In this context results of numerical experiments for a one-dimensional toy problem are provided, showing the robustness of different approaches and empirically verifying theoretical guarantees. Another focus of this review is the exploration of the subspace of data-consistent solutions through explicit guidance to satisfy specific semantic or textural properties.}

\keywords{Inverse Problems; Machine Learning; Robustness; Explorability;}
\maketitle

\section{Introduction}
The goal of image reconstruction is to recover an unknown image from indirect or distorted measurements, i.e., to recover a ground truth image $u$ from measurements 
\begin{equation}
f = A (u) + n.
\label{eq:gen_forward}
\end{equation}
for a forward operator $f$ and (additive) noise $n$. When the forward measurement process is linear,  recovering $u$ becomes a linear inverse problem, which is what we focus on in this paper. Simple approaches compute reconstructions $u$ for \eqref{eq:gen_forward} linearly via least-squares estimates, possibly including an additional smoothing or regularization. Examples of this approach include filtered back projection \cite{Feldkamp:84} for tomographic image reconstruction, and Wiener filtering for image deconvolution, which incorporates regularization through linear filtering in Fourier space. Variational approaches (c.f.~\cite{Benning2018}) to such problems find the minimizer of a suitable cost function, typically consisting of a data fidelity term $E(A,u,f)$ that measures the discrepancy from the observation model \eqref{eq:gen_forward} and a regularizer $R(u)$ that incorporates prior knowledge about the image to be recovered,
\begin{equation}
 \hat{u} = \arg \min_u E(A,u,f) + R(u).
\label{eq:variational}
\end{equation}
At least in finite dimensions, the above perspective relates to the Bayesian approach to inverse problems (see e.g. \cite{dashti2017bayesian}), via the concept of \textit{maximum a-posteriori probability} (MAP) estimates, if the regularizer (with regularization strength $\alpha$) in the form $\exp(-\alpha R(u))$ is integrable w.r.t. $u$. By modeling the unknown image $u$ and the measurements $f$ as realizations of random variables with respective distributions $p(u)$ and $p(f)$, and computing the MAP estimate as the argument that maximizes the posterior distribution $p(u|f)$, the application of Bayes law yields \eqref{eq:variational} with $E(A,u,f) = -\log p(f|u)$ and $R(u) = - \log(p(u))$. While \eqref{eq:variational} is a point estimate, the Bayesian approach to inverse problems inverse models or learns the entire posterior probability $p(u|f)$ and/or sampling schemes for it. Though their notion of solutions is different, both approaches consider the inverse problem to be well-posed if a unique solution exists and depends on the data continuously. 

Classical approaches have thoroughly analyzed ill-posed problems and a large body of work provides stability and convergence guarantees, e.g. by selecting (noise-level-dependent) regularizers with suitable properties in \eqref{eq:variational}. Yet, these regularizers are typically not expressive enough to model the distribution of natural/realistic images faithfully. In the past decade, deep learning has achieved remarkable success in image reconstruction through the ability to capture data-dependent structures, showing great improvements in reconstruction quality over earlier classical approaches. This survey attempts to present an overview of recent deep learning-based image reconstruction methods and discuss issues related to the stability, robustness, and explorability of solutions. 

\textbf{Relation with other review papers:
} There are several existing review articles on classical and machine-learning methods to inverse problems in imaging. These articles range from extensive, general overviews which also introduce basic knowledge and applications (e.g. \cite{arridge2019solving}), to works focussing mostly on proved guarantees in general (e.g. \cite{scarlett2022theoretical}), or for example, convergence guarantees specifically (\cite{10004773}). Other, similar articles give general overviews with stronger foci on some aspects, such as the amount of knowledge available about the forward operator (\cite{ongie2020deep}), or neural network architectures (e.g. \cite{bai2020deep}). In comparison with the existing works, the present article focuses on aspects related to the robustness of point estimators to inverse imaging problems. In this context, we empirically evaluate the robustness of different approaches and verify theoretical guarantees for a one-dimensional toy problem. Another unique focus of this review is the exploration of the subspace of data-consistent solutions through explicit guidance to satisfy specific semantic or textural properties.

\section{Overview of Deep Learning for Inverse Problems in Imaging}
In this section, we give a brief overview of different strategies to solve inverse problems by involving concepts (i.e. learning or parametrization strategies) from deep learning. We distinguish two subcategories, i.e., point estimators (Sec. \ref{sec:point_estimates}) that deterministically predict a single solution to an inverse problem (similar to the MAP estimate \eqref{eq:variational}), and methods that allow stochastic sampling from the posterior $p(u|f)$ (Sec. \ref{sec:distribution_estimates}). We'd like to emphasize however that this distinction is not absolute, as there exist hybrid methods.

\subsection{Deep Learning for Point Estimates to Inverse Problems}
\label{sec:point_estimates}

\paragraph{Direct Approaches} 
In our summary below, we distinguish two types of approaches: Direct approaches (this paragraph) that aim at directly predicting a solution of an inverse problem and are therefore specific to a (class of) forward operators $A$, and network priors that learn (encodings of) the prior probability $p(u)$ only to keep the versatility of variational approaches \eqref{eq:variational}.
\newline
\textbf{Fully Learned Approaches:} A straightforward approach is to train a deep network $\net_\theta$, i.e., a function parameterized by $\theta$, in a supervised manner to directly invert the measurement process: 
\begin{equation}
 \hat{u} = \net_\theta(f), 
\label{eq:network}
\end{equation}
where $\net_\theta$ is trained by minimizing the expectation of some distortion measure $\mathcal{L}$ between the network output and the ground truth over a set of training examples. Common networks are trained using
 simple pixel-wise $\ell_2$ or $\ell_1$ reconstruction losses
\cite{Xie2012,Burger2012,Kim2016,zhu2018image}, losses based on structural similarity metrics \cite{zhao2016loss}, perceptual losses \cite{johnson2016perceptual} or adversarial losses by discriminating between the network output and the clean data distribution \cite{Wang2018,kupyn2019deblurgan} in addition to pixel-wise losses. Networks trained using such an approach have achieved better reconstruction quality in many imaging tasks than classical approaches, as learning on specific datasets allows the networks to capture a data-dependent context.
As such approaches typically do not explicitly take the forward operator $A$ into account in the reconstruction process, they can also be applied when the forward model is not known or cannot be modeled accurately, for example, in blind image restoration tasks \cite{noroozi2017motion,Zamir2021Restormer}. 
The training of such networks usually requires a huge amount of paired training data, which implicitly incorporates the forward operator $A$.
Yet, such fully learned approaches are specific to the task they have been trained on, which contrasts the rather versatile variational approaches \eqref{eq:variational}, in which the forward operator, discrepancy measure, and regularization term can be exchanged flexibly.  Moreover, using deep, rather general parametrizations without encoding knowledge of the measurement process \eqref{eq:gen_forward} explicitly commonly leads to networks whose properties remain poorly understood and are frequently referred to as "black-box" approaches. The latter has led to significant research on combinations of interpretable classical and powerful learning-based approaches.

\noindent\textbf{Learned Post-processors:} The simplest approach to incorporate model knowledge into a learning-based approach is to train a post-processor network on removing artifacts from an initial reconstruction obtained from an analytical linear reconstruction operator $A^\dagger_{\text{reg}}$ such as the adjoint, the pseudo-inverse, or a regularized version thereof (c.f. \cite{mousavi2015deep,7950457,chen2017low}):
\begin{equation}
\hat{u} = \net_\theta(A^\dagger_{\text{reg}}(f)). 
\label{eq:net_post-process}
\end{equation}
A common approach for such post-processing networks is to use residual learning \cite{he2016deep} to recover the difference between initial reconstruction and ground truth. To ensure measurement consistency of solutions obtained by such a post-processing scheme, \cite{schwab2019deep} explicitly constrain the learned residual to be in the null space of the forward operator.  

\noindent\textbf{Unrolled Optimization:} Unrolled optimization \cite{gregor2010learning} uses the knowledge of the forward model to alternate between measurement and image spaces in an iterative algorithm with a fixed number of iterations, where some of the intermediate operations are learned using parameterized deep network modules. Starting from \cite{gregor2010learning}, several works proposed unfolding different iterative model-based algorithms, for instance, learned ISTA \cite{gregor2010learning,Zhang_2018_CVPR}, learned ADMM \cite{sun2016deep}, learned gradient descent \cite{adler2017solving,gong2020learning}, learned primal-dual \cite{adler2018learned}, or proximal gradient algorithms \cite{mardani2018neural,putzky2017recurrent}. Instead of learning a different set of network parameters for the proximal step in each iteration, a few works~\cite{he2018optimizing,gupta2018cnn,aggarwal2018modl} share the same network for the proximal steps across iterations.
Variational networks, as introduced in \cite{kobler2017variational}, are an important subset of unrolled optimization schemes. Here, a proximal gradient descent scheme is incrementally unrolled, with weights at each layer of the network learning a single component of an energy term corresponding to the data fidelity at the corresponding unrolled step. The learned weights are parameters of the data fidelity terms.
This approach found uses such as for MRI reconstruction, see e.g. \cite{hammernik2018learning}.  In comparison with the fully learned approaches, unrolled networks tend to require less training data, and allow for more interpretable, and parameter-efficient learning \cite{Monga2021}. As the unrolled networks are trained typically using a small number of unrolled steps, the inference is also faster in comparison to classical variational approaches, which may need more iterations to converge. On the other hand, testing unrolled networks using more inference steps than used in training typically results in severe artifacts. Recent work \cite{gilton2021deep} addresses this shortcoming through deep equilibrium models \cite{bai2019deep} which incorporate fixed-point convergence by construction. Such models automatically share the same set of parameters across any number of iterations for solving the fixed-point equation.

\paragraph{Neural Network Priors} 
An alternate approach for incorporating model knowledge into learning-based techniques is to use neural networks as priors in variational inference. In contrast to the dedicatedly trained networks, this approach endows the algorithm with the flexibility to handle different measurement models, while improving the performance of handcrafted priors. This class of methods includes learning regularizers, using trained networks such as denoisers, generative models, and even untrained neural networks as priors in variational image recovery. 
\newline
\textbf{Learned Regularizers: }
Classical approaches to learning regularizers include the Field of Experts \cite{FieldOfExperts}, dictionary learning \cite{Aharon06,Mairal08}, learning regularizer via bi-level optimization, e.g. \cite{kunisch13,Chen14}, or learning a (componentwise) proximal operator \cite{Schmidt14}. Learning deep network regularizers often involves explicitly parameterizing the regularization functional $R(\cdot)$ in \eqref{eq:variational} using a neural network $R_\theta$ \cite{li2020nett,ar_nips, uar_neurips2021,kobler2020total,2023goujonconvex} which may be trained based on different objectives.
While \cite{li2020nett} uses a neural network trained to penalize artifacts in the recovered solution, \cite{kobler2020total} trains a neural network regularizer motivated by sparsity penalties. \cite{ar_nips,prost2021learning, uar_neurips2021} learn regularizers which are trained adversarially to distinguish between samples from the training data distribution and degraded samples. Instead of directly parameterizing the regularizers, \cite{rick2017one} learns a proximal operator of a regularizer, and \cite{heaton2022wasserstein} learns the projection operators onto clean data manifolds.
While learned regularizers improve reconstruction performance over handcrafted priors, they may not always guarantee stability or convergence, which requires imposing additional constraints on the regularizer.  \cite{moeller2019controlling} instead train networks to output descent direction with a provable convergence to a minimizer of the (differentiable) energy, while \cite{hannah2022nonsmooth} expanded this approach for non-smooth energies.
Some works constrain the regularizer to ensure a convergent iterative scheme.  \cite{ar_nips, uar_neurips2021} impose Lipschitz-continuity on the regularizer via a  soft-penalty, and \cite{acr_arxiv} enforce convexity of regularizer using input convex networks \cite{amos2017input} for convergence. We refer to \cite{10004773} for a review of learned reconstruction methods with convergence guarantees.
As mentioned previously, training supervised neural networks takes huge amounts of labeled data. A common strategy to circumvent this restriction is explicitly learning priors on patches of input data, as it allows for huge amounts of training data, extracted from only few complete input samples.\cite{zoran2011learning,altekruger2023patchnr,prost2021learning} model patch-based priors based on Gaussian mixture models, convolutional neural networks or normalizing flows.
It needs to be emphasized that learned regularizers can also be used in from a Bayesian perspective and allow for sampling.
 
\noindent\textbf{Denoisers as Priors:}
Pretrained denoisers have been employed as priors in image recovery- as proximal operators, or in a functional representing the gradient of regularizer.
Plug-and-Play (PnP) methods \cite{venkatakrishnan2013plug,chan2016plug} replace proximal operators of a regularizer by generic denoisers such as non-local means \cite{nonlocalmean} or BM3D \cite{bm3d} in proximal splitting algorithms. Subsequently \cite{Zhang2017b,meinhardt2017learning,zhang2021plug} proposed the use of pretrained neural network denoisers as proximal operators with good empirical results. 
An alternate approach is regularization by denoising (RED)  using denoisers $D_ \theta$  in a regularization functional of the form $\langle u,u-D_\theta(u) \rangle$ \cite{arjomand2017deep,Romano2017}  in a gradient descent based scheme.  While both PnP and RED approaches empirically provide very good reconstructions, they require strong conditions on the denoiser to have convergence guarantees. The denoiser replacing the proximal operator should be non-expansive, or in the  RED framework, the denoiser should additionally have a symmetric Jacobian. These restrictive conditions are not satisfied by arbitrary denoising networks \cite{reehorst2018regularization}. A few approaches constrain the denoiser to satisfy properties required for convergence, for instance, \cite{ryu2019plug,terris2021enhanced}  train denoisers with constrained Lipschitz constants, \cite{cohen2021has}  derive image denoisers with symmetric Jacobians, and \cite{hasannasab2020parseval}  parameterize 1-Lipschitz operators for denoising.
Instead of constraining the denoisers, \cite{sommerhoff2019energy} project the outputs of arbitrary denoisers onto the cone of descent directions to a given energy in a (proximal) gradient descent algorithm for provable convergence.

\noindent\textbf{Untrained Neural Network Prior: }
In \cite{ulyanov2018deep} Ulyanov et al. proposed to use the structure  of a randomly initialized convolutional generator to capture natural image statistics,   referred to as  `Deep Image Prior' (DIP), and used this to solve inverse problems  by optimizing  untrained network weights to minimize reconstruction error:
\begin{equation}
 \hat{u} = \net(z_0;\hat{\theta}) \text{ ~s.t.~ } \hat{\theta}=\underset{\theta}{\arg \min}~\|f-A\net(z_0;\theta)\|^2.
\label{eq:DIP}
\end{equation}
Their work used an over-parameterized UNet \cite{Ronneberger2015} for $\net$ and suggested early stopping of the optimization in \eqref{eq:DIP} to prevent overfitting. \cite{heckel2019deep} instead use an under-parameterized non-convolutional generator which prevents overfitting. More recent works \cite{chen2020dip,ho2021neural,liu2023devil} even search for neural architectures to be used as deep image priors.  \cite{jagatap2019algorithmic} present a projected gradient descent scheme for solving \eqref{eq:DIP} using under-parameterized networks \cite{heckel2019deep} and provide convergence guarantees for their scheme. 
A few works \cite{cheng2019bayesian,liu2019image,mataev2019deepred} have also combined deep image priors with additional regularization. \cite{cheng2019bayesian} considers a Bayesian perspective of the deep image prior as a Gaussian process and computes MMSE estimate $\hat{u}$ by optimizing both $\theta$ and $z$ using $\ell_2$ regularization on both. \cite{liu2019image} employ TV regularization on the DIP network output. \cite{mataev2019deepred} use a combination of deep image prior and regularization by denoising, and \cite{van2018compressed} combine DIP with learned regularization. More theoretical works like \cite{dittmer2020regularization} show that under specific conditions guarantees regarding regularity, convergence with regard to the noise level, existence of solution, and stability can be given. Similarly \cite{habring2022a} shows that untrained network priors with specific architectures and parameter/activation function choices allow interpretations as Landweber iterations or unrolled proximal gradient descent, with further analytical analysis.
DIP and similar approaches can also be of use in Bayesian approaches, see e.g. \cite{laumont2022bayesian}.

\subsection{Deep Learning for Bayesian Approaches to Inverse Problems}
\label{sec:distribution_estimates}
While point estimators are certainly interesting, they do not describe all possible solutions given the observation $f$, which are commonly represented by the \textit{posterior} $p(u|f)$. Stochastic approaches to inverse problems allow to sample solutions from $p(u|f)$, which is particularly desirable for under-determined inverse problems. On the theoretical side works like \cite{stuart2010inverse} give an overview of important theoretical foundations of Bayesian approaches to inverse problems, while \cite{sprungk2020on}  focus on stability and show that under mild assumptions Bayesian inverse problems can even be well-posed.
This is useful for uncertainty quantification \cite{narnhofer2022posterior,siahkoohi2020faster} as it not only allows computing point estimates such as the minimum mean squared error (MMSE) estimator but also deriving higher-order statistics such as the variance of each pixel of the reconstructions. Existing deep learning approaches to stochastic image recovery frequently utilize conditional or unconditional generative models. The former approach involves training a conditional generative model, i.e., a model that gets additional inputs such as the measured data, for a specific recovery task, and the latter approach uses a pre-trained generative model as a prior for the image recovery. We will briefly summarize works in both directions below.

\noindent\textbf{Conditional Generative Models:} 
 Conditional generative models train generative approaches with the data of an inverse problem as an additional input, where the training phase ensures, or at least encourages data-consistency of the predicted solutions following \eqref{eq:gen_forward}. Consequently, they can sample multiple solutions $\hat{u}$ for a given observation $f$  \cite{ardizzone2018analyzing,padmanabha2021solving,li2022srdiff,lugmayr2020srflow,peng2020generating}.  Most commonly this is achieved by supervised training of conditional generative models such as conditional generative adversarial networks (GANs) \cite{jo2021tackling,bahat2020explorable,ohayon2021high}, conditional flow models  \cite{winkler2020learning,lugmayr2020srflow,siahkoohi2020faster,Jo_2021_CVPR}, or conditional diffusion models \cite{saharia2021image,li2022srdiff,saharia2022palette,Whang_2022_CVPR}. A few of these methods also guarantee consistency of the reconstruction with input either by an explicit projection operation \cite{bahat2020explorable}, or by choosing inherently invertible generative models \cite{lugmayr2020srflow,Jo_2021_CVPR,ardizzone2018analyzing,padmanabha2021solving}. Only a few works \cite{sim2020optimal,Sun_Bouman_2021,altekruger2023wppnets,runkel2023learning,rezende2015variational} use unsupervised or unpaired learning to learn conditional generative models. Any of the above approaches represents the posterior as a transformation of a simple distribution (e.g. a Gaussian) via a parameterized mapping, the conditional generative model, such that samples from the posterior can be drawn by feeding different samples from the initial distribution into the trained network, or, more formally, considering the push-forward of the latent distribution under the conditional generative model.

\begin{figure}[thb]
    \centering
    \includegraphics[clip, trim=3cm 1cm 3cm 0cm,width=0.9\textwidth]{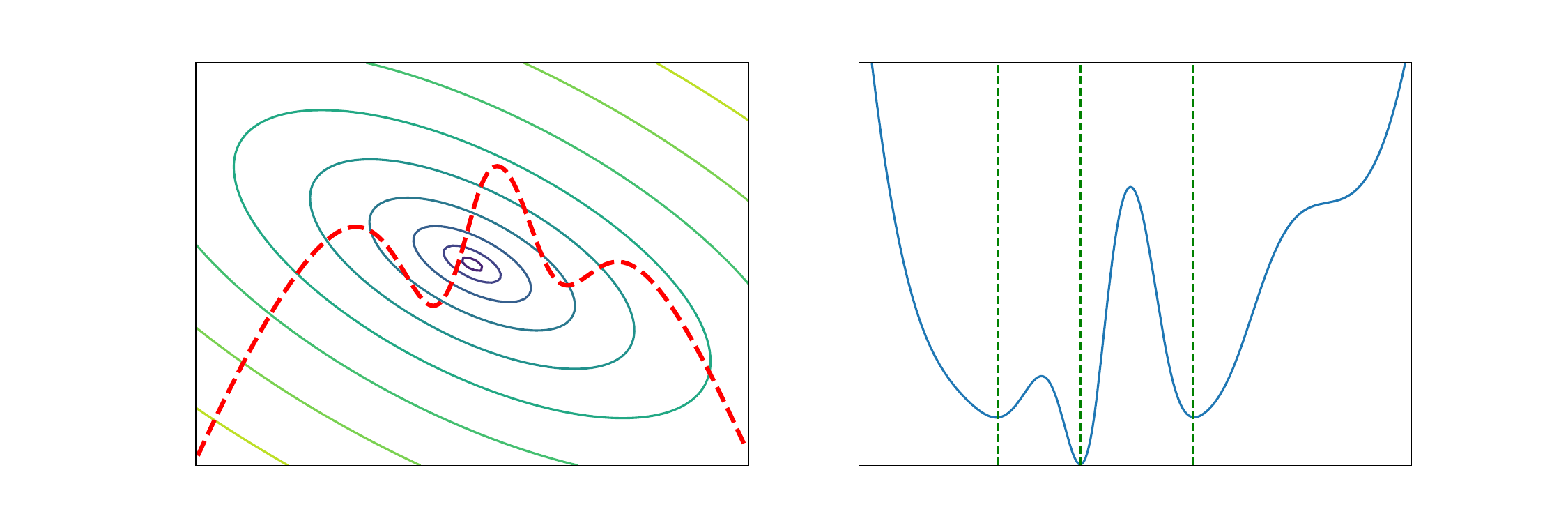}
    \caption{The approach \eqref{eq:ganprior} corresponds to a posterior that is the product of the likelihood $p(f|u)$ and a prior $p(u)$ that restricts $u$ to the range of a generator $\net_\theta(\hat{z})$. The left plot illustrates level lines of $-\log(p(f|u))$ in 2d along with a lower dimensional manifold that is the range of $\net_\theta(\hat{z})$ as a dashed red line. The resulting costs are shown on the right. Running gradient descent from different starting points can merely sample local minima (dashed green lines) that correspond to local maxima of the posterior.}
    \label{fig:initialization}
\end{figure}

\noindent\textbf{Generative Priors: } Generative models such as generative adversarial networks (GANs) \cite{goodfellow2014generative},   variational autoencoders (VAEs) \cite{kingma2013auto},  normalizing flows \cite{dinh2017density}, and diffusion-based or score-based models \cite{song2020score,ho2020denoising} are trained to produce new samples from the underlying distribution of the training data, and therefore can serve as useful priors when the image to be recovered belongs to this distribution. These models learn a generator $\net_\theta$ to transform a simple distribution $p(z)$ on a latent space (e.g. a Gaussian) to the image distribution $p(u)$ (as opposed to the previous paragraph, where models try to directly predict the posterior). \cite{bora2017compressed}  proposed the use of deep generative model priors for image recovery by optimizing for a vector in the smaller dimensional latent space of a trained GAN or a VAE to minimize the reconstruction error: 
\begin{equation}
 \hat{u} = \net_\theta(\hat{z}) \text{ ~s.t.~ } \hat{z}=\underset{z}{\arg \min}~\|f-A\net_\theta(z)\|^2,
\label{eq:ganprior}
\end{equation}
with an $\ell_2$ regularization on $z$ using simple gradient descent-based methods, and demonstrated significant improvements over the classical priors for compressive sensing with a small number of measurements. For compressed sensing using random Gaussian matrices, they show that \eqref{eq:ganprior} results in solutions close to the ground truth with high probability under certain conditions. Their work was later extended to non-linear inverse problems in \cite{hand2018phase,nonlinearbayesian}. \cite{Yeh_2017_CVPR} proposed image inpainting by using Poisson blending using the image that is closest in the latent space of the generator to the input corrupted image. \cite{chandramouli2022generative} proposed latent space optimization of a generative autoencoder for light field recovery.
\cite{shah2018solving, Raj_2019_ICCV}  investigated the use of projected gradient descent, and \cite{latorre2019fast} proposed the use of the alternating direction method of multipliers (ADMM) for image recovery using GAN priors.  \cite{prost2023inverse}  utilize hierarchical VAEs in an efficient Plug-and-Play algorithm for general inverse problems. 
An advantage of latent space optimization is the ability to obtain multiple solutions by using different initial latent codes \cite{menon2020pulse,marinescu2021bayesian,pan2021exploiting},  which can be accelerated by finding latent space directions in the null space of the forward operator \cite{explore_solution_gan_latent}. Yet, such strategies rather sample different local maxima of the posterior $p(u|f)$ than reflecting the posterior itself, see Fig.~\ref{fig:initialization} for an illustration.
A major limitation of the latent space optimization \eqref{eq:ganprior} is that samples outside the range manifold of the generator cannot be reconstructed accurately resulting in a non-trivial representation error. Subsequent works attempt to reduce this representation error using different approaches. \cite{dhar2018modeling} allows a small deviation of the recovered image from the range of a generator with sparsity prior on their difference, which is extended to optimizing intermediate layer representations in \cite{daras2021intermediate}. \cite{hussein2020image,pan2021exploiting} adopt a two-step approach of latent space optimization followed by fine-tuning both the latent vector and generator parameters.  \cite{doi:10.1137/21M140225X} proposes a framework for inverse problems using VAEs by considering a joint posterior distribution of latent $z$ and image space $u$ which guarantees convergence to a stationary point. 
\cite{asim2020invertible} replace the GAN prior in \eqref{eq:ganprior} with a flow-based generator, with an $\ell_2$ regularization on $z$. \cite{pmlr-v139-whang21a,pmlr-v161-kothari21a} generalize this to
arbitrary differentiable measurement operators and measurement noises using a maximum a-posteriori framework, with \cite{pmlr-v161-kothari21a} using a generalized version of flow models which progressively increase dimension from a low-dimensional latent space. Works like \cite{alberti2022continuous} investigate generative models with continuous outputs, focussing on theoretical aspects, and show that under certain conditions generative priors can be injective, allowing Lipschitz-Bounds for their usage in inverse problems.

Energy based models \cite{du2019implicit} are a class of generative methods which learn the prior directly by learning a neural network representing an energy functional which assigns low energy values to samples in the distribution of training data  and high energy values otherwise through maximum-likelihood training. Energy based models have also been used for inverse imaging tasks such as MRI reconstruction \cite{10237244}, and computed tomography \cite{zach2022computed}.
 The works \cite{jalal2021robust,kadkhodaie2021stochastic,laumont2022bayesian,jalal2021instance} adopt Langevin dynamics for linear inverse problems and incorporate the guidance from measurement through the gradient of the log-posterior into their iterative process, or via a projection operation \cite{kadkhodaie2021stochastic}.
Several recent works incorporate the knowledge of the forward operator to modify the reverse sampling process in denoising diffusion models.  This can be done by alternating between a standard reverse diffusion step and a projection operation for promoting measurement consistency \cite{choi2021ilvr,chung2022come,wang2022zero,lugmayr2022repaint}. An alternate approach is to e.g. use a gradient descent step on the data fidelity term \cite{chung2022diffusion,chung2022improving} or the pseudo-inverse of the forward operator \cite{song2023pseudoinverseguided} using a clean estimate at each step of the reverse diffusion process, where \cite{chung2022improving} additionally include a correction step through projection. While approaches that employ only one projection operation per denoising step (such as \cite{wang2022zero,lugmayr2022repaint}) are faster, they are restricted to linear inverse problems.
On the other hand, guidance using the gradient of a data fidelity term \cite{chung2022diffusion} can be applied even to non-linear inverse problems, yet it is more expensive as it requires back-propagation through the diffusion model weights at each iteration.  
\cite{whang2021composing,Feng_2023_ICCV} train conditional flow based models to parameterize the posterior $p(u|f)$, given a pretrained generative model representing the prior. More recently, \cite{mardani2023variational} adopt diffusion models in a regularization-by-denoising framework, and \cite{zhu2023denoising} demonstrate their utility for plug-and-play image restoration as an effective alternative to the standard Gaussian denoisers.

\section{Stability and Robustness}
As deep learning approaches are increasingly adopted in image recovery tasks, characterizing the vulnerabilities and instabilities of neural networks for image recovery is important, especially in safety-critical applications like medical imaging. While adversarial robustness is extensively studied for image classification, see e.g. \cite{szegedy2013intriguing,goodfellow2014explaining,aleksandermadry}, it is less studied in the context of image recovery.
The notion of robustness itself is very different for classification and reconstruction problems. For a classifier, robustness can be characterized by the minimal perturbation which can cause a sample to cross the decision boundary, leading to a change in classification outcome. 
For reconstruction tasks, the outputs are not discrete labels, and there is no notion of a decision boundary. Instead, the output of the reconstruction algorithm should vary continuously/smoothly with changing inputs. 
The latter links to the mathematical study of inverse problems in infinite-dimensional spaces, where the pseudo-inverse of the linear operator $A$ in \eqref{eq:gen_forward} is neither continuous nor defined everywhere as soon as $A$ is compact and has an infinitely dimensional range. Consequently, one has to balance the desire to have a suitable type of continuity in the reconstruction with the faithful approximation of the pseudo-inverse depending on the expected noise. Furthermore, the ill-posedness of the inverse problem might arise from a lack of uniqueness, e.g., due to a forward operator with a non-trivial null-space.    
  
For finite dimensional linear inverse problems, the degree of "ill-posedness" can be quantified by the condition number $\kappa(A) = \frac{\sigma_{max(A)}}{\sigma_{min(A)}}$ of the given operator $A$, with $\sigma_{\{max,min\}}(A)$ denoting the largest/smallest singular value. In the infinite-dimensional setting, $0$ is an accumulation point of the singular values and the severity of the ill-posedness is characterized by how fast the singular values converge to zero. 

The aforementioned notion of ill-posedness motivates the (classical) linear regularization of the singular values (spectral regularization): Consider a forward operator $A$ with singular value decomposition 
\begin{equation}
\label{eq:svd}
    Au = \sum_{i=1}^\infty\sigma_i\langle u,\mu_i\rangle\nu_i,
\end{equation}
where $\sigma_i$ again denote the singular values, while $\mu_i$ and $\nu_i$ are the $i$-th left/right singular vectors. One replaces the (unbounded) $1/\sigma_i$ that arise in the pseudoinverse by a suitable bounded approximation $g_{\alpha_i}(\sigma_i)$, i.e.,
\begin{equation}
\label{eq:spectral regularizer}
\begin{aligned}
    R_\alpha(f) = \sum_{i=1}^\infty g_{\alpha_i}(\sigma_i)\langle f,\nu_i\rangle\mu_i
\end{aligned}
\end{equation}
with regularization functions $g_{\alpha_i}(\sigma)$ parameterized by one or multiple $\alpha_i$ that determine a (noise-dependent) balance between the boundedness of the reconstruction operator and the faithfulness of approximating $A^\dagger$. Although being limited to specific linear regularizations (and therefore typically being suboptimal in imaging applications) the analysis of such approaches is well-established and conditions for choosing $\alpha$ such that $R_\alpha(f)$ converges to the true solution are well-established. Recent work has extended such analysis to learned regularization functions $g_{\alpha_i}$. It demonstrated that an analytical solution for the optimal regularization can be computed and yields stability guarantees, see e.g. \cite{bauermeister2020learning,kabri2022convergent}.

A common way to characterize stability in finite dimensions, particularly in the neural network community, has been to use the notion of Lipschitz continuity. If a reconstruction algorithm $\net$ satisfies
\begin{equation}
\label{eq:lipschitz}
    \|\net(f+\delta)-\net(f)\| \leq L\|\delta\|,
\end{equation}
then $\net$ is a Lipschitz continuous mapping with Lipschitz constant $L$, where $\| \cdot \|$ on both sides of \eqref{eq:lipschitz} is commonly chosen to be the $\ell_2$ norm. From the point of view of stability, a small value of $L$ is desirable to ensure that the maximal change in the reconstruction caused by a small change in measurements remains small.  
While Lipschitz continuity provides a useful notion of stability, analyzing the stability of common neural networks in terms of Lipschitz constants is difficult, owing to the high complexity involved in its exact computation, even for moderately sized neural networks \cite{NEURIPS2020_5227fa9a}. In particular, computing the smallest Lipschitz constant was shown to be NP-hard even for a 2-layer fully connected network in \cite{virmaux2018lipschitz}.  As a result most works \cite{weng2018towards,virmaux2018lipschitz,combettes2020lipschitz} only compute approximations and upper bounds for the smallest Lipschitz constant of neural networks. 
On the other hand, stability in terms of a Lipschitz continuous network seems to come at the cost of the reconstruction performance. For instance, \cite{sommerhoff2019energy} observed that enforcing non-expansiveness ($L\leq1$) drastically decreased the denoising performance of neural network denoisers. Moreover, even analytically, for an ill-posed problem with a forward operator $A$ that yields a one-to-one correspondence between ground truth and measurements, the Lipschitz constant has to be noise-level-dependent and has to tend to infinity as the reconstruction operator approximates $A^\dagger$. A noise-level independent, fixed (small) value of $L$ implies the inability of $\net$ to accurately reconstruct the ground truth.
See \cite{gottschling2020troublesome} for a discussion. 

\cite{10015061}  show that variational energy minimization approaches  of the specific form  $
   \hat{u}=\arg\min_u\|Au-f\|^2+\lambda\|u\|^p_p\text{ for }  p\in(1,\infty)$
 show good stability properties, with Tikhonov $\ell_2$ regularized reconstruction map being globally Lipschitz continuous. The
reconstruction map is locally Lipschitz continuous in the measurement space for $ p\in(1,2)$,  and globally $\frac{1}{p-1}$ Hölder continuous\footnote{$\net$ is $\alpha$ Hölder continuous if   $\|\net(f+\delta)-\net(f)\|_p \leq K\|\delta\|_2^\alpha$}  for $p\in (2, \infty)$. In general, all variational energy minimization approaches permit a stability estimate in the case of linear inverse problems using convex regularizers, as shown in \cite{burger2007error}. The optimality condition for \eqref{eq:variational} with $E(A,u,f) = \frac{1}{2}\|Au-f\|^2$ is
\begin{equation}
\label{eq:optimality_linear_variational}
0 \in A^*(Au-f)+\partial R(u).
\end{equation}
The $\partial$ symbol in \eqref{eq:optimality_linear_variational} denotes the subdifferential, defined for a convex, real-valued function $f$ via $f(x)\leq f(x_0) + \langle a, x-x_0\rangle$, where $a$ is a valid subgradient of the function at $x_0$ and with the set of all $a$ being called the subdifferential. Intuitively speaking the subdifferential of a function at a point is the set of gradients of all hyperplanes lying completely below the graph of the function.
Taking the difference between the two optimality conditions arising from two different measurements $f_1$ and $f_2$ with their corresponding reconstructions $u_1$ and $u_2$ and subsequently taking the inner product with $u_1 - u_2$, we find that there have to exist subgradients $p_1 \in \partial R(u_1)$ and $p_2 \in \partial R(u_2)$ such that
\begin{equation}
\label{eq:linear_variational_part1}
\begin{aligned}
    0 & = \langle A^*(Au_1-f_1)-A^*(Au_2-f_2)+p_1-p_2, u_1-u_2\rangle\\
      & = \|Au_1-Au_2\|^2-\langle f_1-f_2, Au_1-Au_2 \rangle+\langle p_1-p_2, u_1-u_2 \rangle.
\end{aligned}
\end{equation}
The second term can now be bounded from above by applying $\langle a,b\rangle \le \|a\|\|b\|$ and $ab \le \frac{a^2}{2}+\frac{b^2}{2}$:
\begin{equation*}
\langle f_1-f_2,Au_1-Au_2\rangle \le \frac{1}{2}\|f_1-f_2\|^2+\frac{1}{2}\|Au_1-Au_2\|^2.
\end{equation*}
Rewriting this equation yields
\begin{equation*}
\begin{aligned}
    \frac{1}{2}\|f_1-f_2\|^2 \ge \frac{1}{2}\|Au_1-Au_2\|^2+\langle p_1 - p_2, u_1-u_2 \rangle
 \label{eq:stability_tv1}
\end{aligned}
\end{equation*}
or alternatively that
\begin{equation}
\begin{aligned}
    \frac{1}{2}\|f_1-f_2\|^2 \ge \frac{1}{2}\|Au_1-Au_2\|^2+\mathcal{D}_R(u_1,u_2)
 \label{eq:stability_tv}
\end{aligned}
\end{equation}
with $\mathcal{D}_R(a,b)$ denoting the symmetric Bregman distance with respect to the convex regularizer $R$. Note that -- as opposed to \eqref{eq:lipschitz} -- the natural stability of a variational approach is the sum of a data consistency and a regularization-specific measure of distance. The latter can be a rather weak notion of a difference as only the properties $\mathcal{D}_R(u_1,u_2)\geq 0$ and $\mathcal{D}_R(u,u)=0$ can be guaranteed, unless $R$ is $m$-strongly convex, in which case $\mathcal{D}_R(u_1,u_2)\geq m\|u_1 - u_2\|^2$. Therefore, even though learned convex regularizers provably satisfy \eqref{eq:stability_tv}, their underlying symmetric Bregman distance might remain difficult to interpret. Interesting future research could therefore involve architectures or additional loss functions for encouraging a particularly meaningful Bregman distance. We further refer the reader to \cite{benning2011error} for error estimates with non-quadratic data fidelity terms. 

\begin{figure}
    \centering
    \begin{subfigure}[b]{0.2\textwidth}
        \centering
        \includegraphics[width=1.0\textwidth]{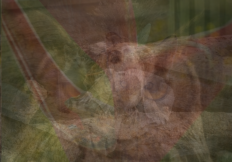}
        \vspace{0.3cm}
        \caption{Mean of the images in (b)}
    \end{subfigure}\hfill
    \begin{subfigure}[b]{0.7\textwidth}
        \centering
        \includegraphics[width=1.0\textwidth]{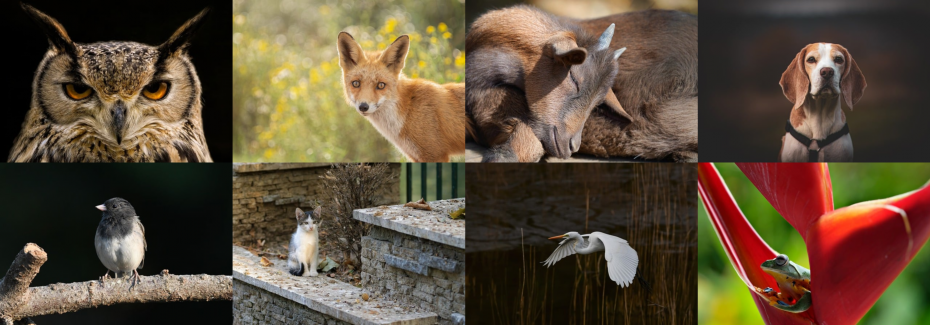}
        \caption{Various natural images}
    \end{subfigure}
    \caption{Convex regularizers rate convex combinations (a) of images (b) as at least as "natural" as one of the images itself. }
    \label{fig:convex_image}
\end{figure}
Yet, although the analysis of convex variational methods, their ability to determine global minimizers and stability properties like \eqref{eq:stability_tv} are highly appealing, they are systematically suboptimal to act as a prior for natural images as illustrated in Fig.~\ref{fig:convex_image}: According to any convex regularizer $R$, any convex combination of natural images is "at least as likely" to be a natural image as one of its constituting images. This property makes a strong case for nonconvex regularizers, which, even though being more mathematically challenging to analyze, are slowly coming into the focus of the research community (see e.g. \cite{pinilla2022improved} for invex regularizers). Since nonconvex or general learning-based approaches, however, remain very challenging to analyze mathematically, a large body of work on the empirical analysis of stability exists for different notions of robustness including robustness to adversarial perturbations, robustness in recovering fine details, robustness to changes in forward measurement operator and robustness to distribution shifts in data. We will discuss these notions of robustness along with recent findings in the following subsections. 

\subsection{Adversarial Robustness}
Adversarial robustness of a learned reconstruction operator $\net_{\theta}$ can be quantified by the maximum deviation caused in the reconstruction by a small perturbation in the measurement. The adversarial perturbation causing the maximum reconstruction can be obtained as
\begin{equation}
    \delta_{\text{adv}} = \underset{\delta \in B_\epsilon}{\arg \max} ~ ~d(\net_{\theta}(f+\delta),\net_{\theta}(f))
    \label{eq: adv_robustness_def}
\end{equation}
for a suitable measure of distance $d$ between two reconstructions, most commonly $d(\net_{\theta}(f+\delta),\net_{\theta}(f))=     
    \left\|\net_{\theta}(f+\delta)-\net_{\theta}(f)\right\|_{2}$, 
and a suitable set $B_\epsilon$ of perturbations, e.g. $B = \{\delta ~|~ \|\delta\|_\infty \leq \epsilon\}$. When the maximum deviation $d(\net_{\theta}(f+\delta),\net_{\theta}(f))$ is small with respect to $\delta_{\text{adv}}$, the reconstruction operator $\net_{\theta}$ can be considered robust. This notion of robustness is closely related to Lipschitz continuity. In practice the optimization problem in \eqref{eq: adv_robustness_def} is seldom solved exactly as this is prohibitively complex, and is approximated using a small number of projected gradient ascent steps. We employ a similar approximation called "Fast Gradient Sign Method" (FGSM) \citep{goodfellow2014explaining}, which simplifies the approximation further by only utilizing the sign of the gradient, in only a single step to obtain $\ell_\infty$ norm constrained adversarial examples.
Adversarial perturbations to maximize reconstruction error with a perturbation budget of $\epsilon$ are generated using the FGSM attack as:
 \begin{equation}
\delta_{adv} = \epsilon\cdot\mathrm{sign}\nabla_{\delta}\|\net_\theta\left(f+\delta\right) -\net_{\theta}(f)\|_2
\end{equation}
Instead of a single step, the projected gradient descent attack (PGD) \cite{aleksandermadry} uses multiple FGSM steps with step size $\alpha$ while clipping the adversarial noise to the perturbation budget at each step which results in a stronger attack than FGSM. Adversarial perturbations to maximize reconstruction error using PGD attack can be generated as:
 \begin{equation}
 \label{eq:linf_pgd}
\delta^{t+1} = \mathrm{clip}\left(\delta^t+\alpha\cdot\mathrm{sign}\left(\nabla_{\delta^t}\|\net_\theta\left(f+\delta^t\right) -\net_{\theta}(f)\|_2\right), [-\epsilon,\epsilon]\right)
\end{equation}
Madry \etal~\citep{aleksandermadry} further propose to run the PGD attack starting from different random points within the allowed set of perturbations to deal with the non-convexity of loss landscape.

Recent works starting from \cite{antun2019instabilities,Choi_2019_ICCV} have characterized the instabilities of deep learning-based image recovery methods. The authors of \cite{Choi_2019_ICCV} demonstrate the susceptibility of deep networks for image super-resolution to adversarial perturbations, with a focus on untargeted attacks.  \cite{antun2019instabilities} show that end-to-end trained deep networks for image recovery are susceptible to adversarial perturbations. They find that perturbations optimized for the networks do not transfer to classical approaches like $\ell_1$ minimization with sparsity constraints, and conclude that these classical methods are more robust than learned approaches. On the other hand, \cite{genzel2020solving,darestani2021measuring,gandikota2023evaluating} analyze the stability of both classical and deep learning approaches to image recovery, and show that even classical approaches are susceptible to adversarial perturbations optimized for these methods, considering $d = \|\cdot \|_2$ and $B_\epsilon$ being the $\ell_\infty$ ball.
Further, the work \cite{gandikota2023evaluating} also shows that adversarial perturbations optimized for classical approaches transfer well to learned approaches,  indicating that these methods do share some common directions of vulnerability. One would anticipate such directions of vulnerability to lie in a subspace spanned by singular vectors of the forward operator corresponding to small singular values, which remains an interesting investigation for future research, particularly considering that classification networks remain susceptible to low-frequency attacks, c.f. \cite{guo20a}, which are not part of the aforementioned subspace for many prominent inverse imaging problems.

Because most works analyze robustness by considering deviations in the reconstructed image in terms of the $\ell_2$ metric, the conclusion that classical (convex, variational) methods are also susceptible to adversarial perturbations in general, would be too broad, particularly considering that they are \textit{provably robust} to $\ell_2$ perturbations in the data space when robustness is measured in terms of data consistency and the symmetric Bregman distance of the regularizer, see \eqref{eq:stability_tv1}. We will illustrate the differences between these perspectives in a toy problem of recovering one-dimensional signals following the empirical evaluation of Genzel \etal \cite{genzel2020solving} below. 

\paragraph{Different Notions of Robustness - Numerical Experiments on a Toy Problem.} We generate signals $u$ as discretizations of piece-wise constant functions with a random number of jumps, each varying in height. The forward operator $A \in \mathbb{R}^{\frac{N}{2}\times N}$ is chosen to be a compressed sensing operator with $N = 1024$, containing random numbers drawn from a Gaussian distribution with mean $0$ and variance $0.05$. 
We compare the model-based approaches of Tikhonov ($\ell_2$-squared) and total variation (TV) reconstruction to learning-based approaches. To show the effects of adversarial attacks on variational approaches we show two versions of Tikhonov, using a very large and the optimal (rather small) regularization strength. For TV we choose the optimal regularization strength as determined by a hyperparameter search, to show the effect in the optimal, practically relevant case. The learning-based approaches include a post-processing U-Net \cite{Ronneberger2015} architecture, an end-to-end learned Tiramisu architecture \cite{jegou2017one} without any model based components and a model-motivated (plug-and-play) architecture (denoted as \textit{ItNet}), incorporating a U-Net based proximal step.
We measure their robustness against adversarial attacks by projected gradient descent with a projection to an $\ell_\infty$-ball with radius $\epsilon$. We chose $\epsilon = 0.2$ in our experiments.

\begin{figure}
    \centering
    \begin{subfigure}[b]{0.4\textwidth}
        \centering
        \includegraphics[width=1\textwidth]{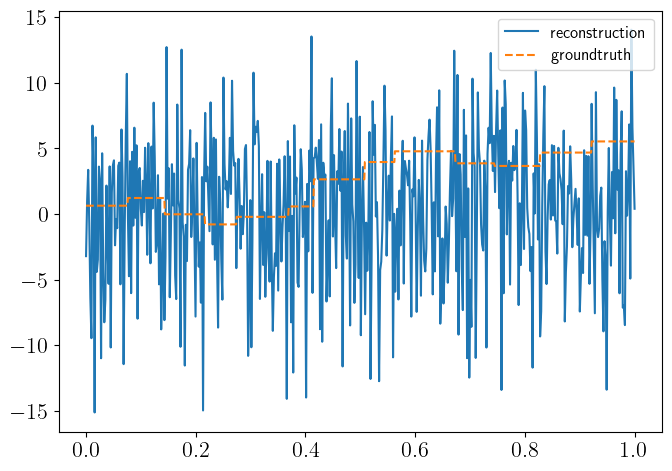}
        \caption{Original measurement.\\\ }
    \end{subfigure}
    \begin{subfigure}[b]{0.4\textwidth}
        \centering
        \includegraphics[width=1\textwidth]{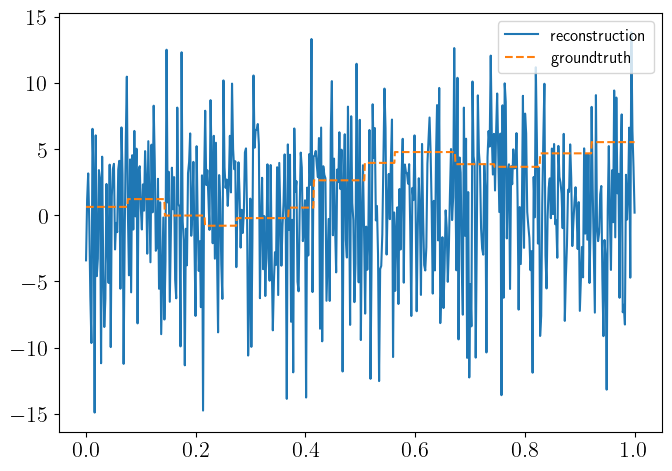}
        \caption{Measurement including adversarial perturbations, here for the highly regularized Tikhonov case.}
    \end{subfigure}
    \caption{The measurement (and its adversarial version)  we attempt to reconstruct in our experiments. As is obvious, the difference between both is negligible.}
    \label{fig:measurements}
\end{figure}

Our experiments consist of a compressed sensing task (as done in \cite{genzel2020solving}), where we aim to restore an (originally sparse) signal from measurements taken by an underdetermined matrix with i.i.d entries, sampled from a Gaussian distribution. The measurement (and an adversarial perturbation) are shown in figure \ref{fig:measurements}.

Details about the optimization process, hyperparameter choices, and further information can be found in the codebase, which will be made publically available\footnote{\url{https://github.com/AlexanderAuras/GAMM-Overview-23/}}.
In our first experiment, we directly reconstruct the signal by employing a Tikhonov-like regularized inversion method 
\begin{align}
\label{eq:tikhonov}
    \hat{u}_{Tik} &= (A^T A + \alpha D^T D)^{-1} A^T f
\end{align}
where the matrix $D$ serves as a finite difference matrix, $\hat{u}_{Tik}$ is the estimated solution, and $\alpha$ is a parameter that balances the fidelity to the data with the smoothness of the reconstruction.

\begin{figure}
    \centering
    \begin{subfigure}[b]{0.4\textwidth}
        \centering
        \includegraphics[width=1\textwidth]{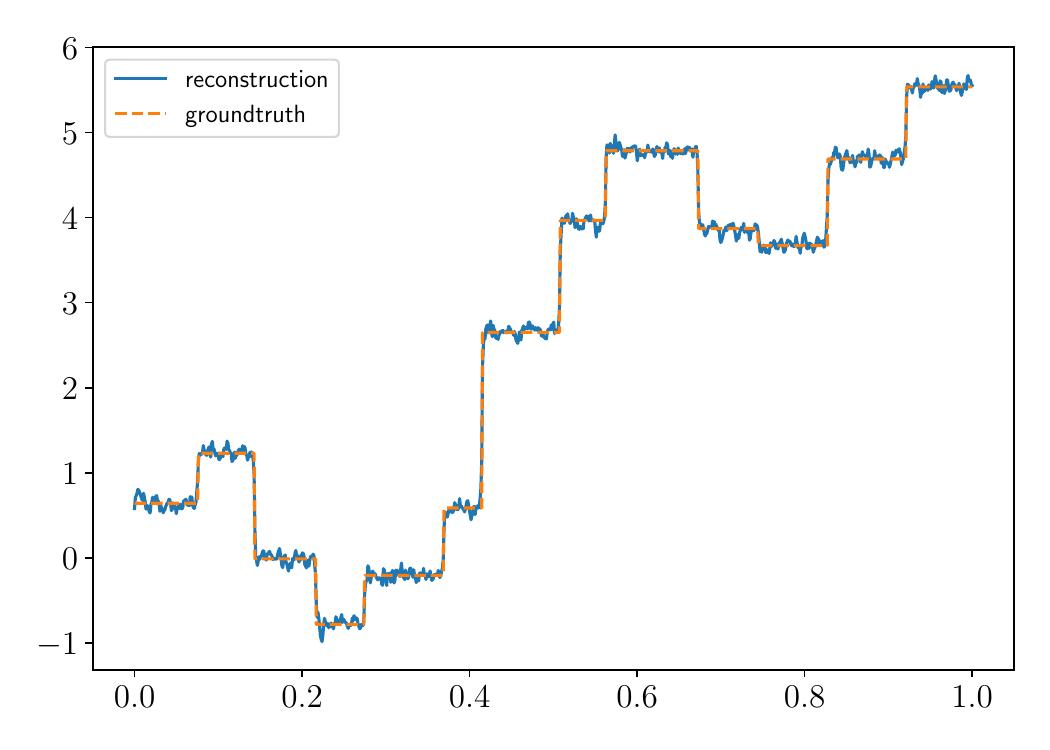}
        \caption{Tikhonov reconstruction with $\alpha = 10^{-7}$}
    \end{subfigure}
    \begin{subfigure}[b]{0.4\textwidth}
        \centering
        \includegraphics[width=1\textwidth]{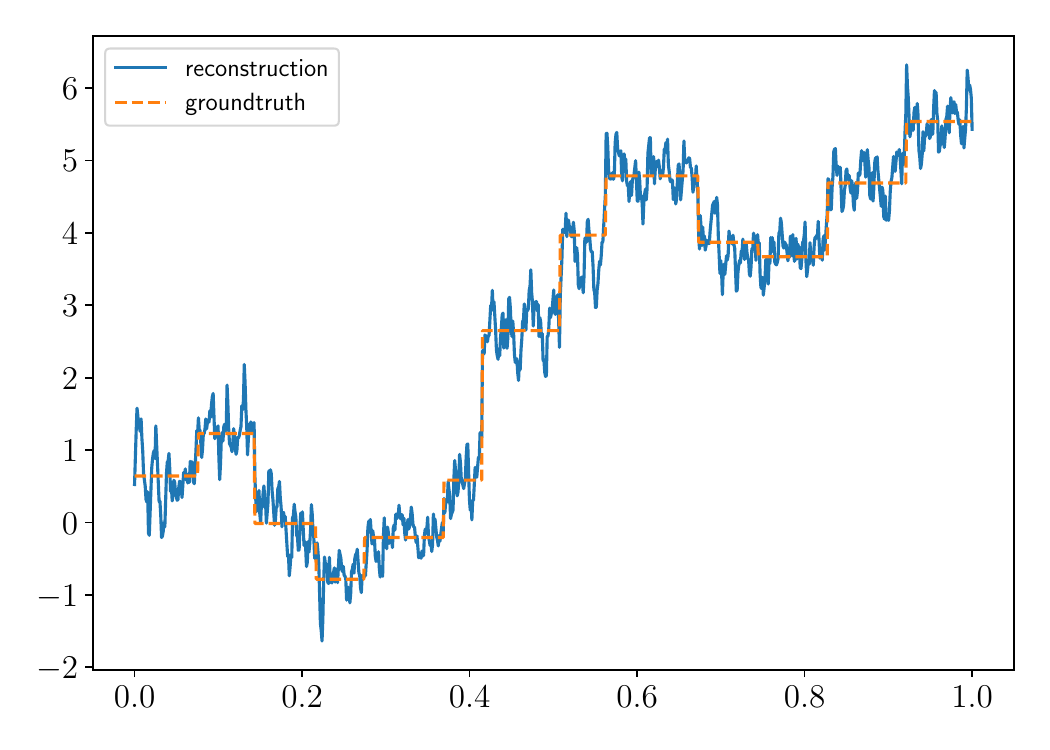}
        \caption{Tikhonov adversarial reconstruction with $\alpha = 10^{-7}$}
    \end{subfigure}
    \begin{subfigure}[b]{0.4\textwidth}
        \centering
        \includegraphics[width=1\textwidth]{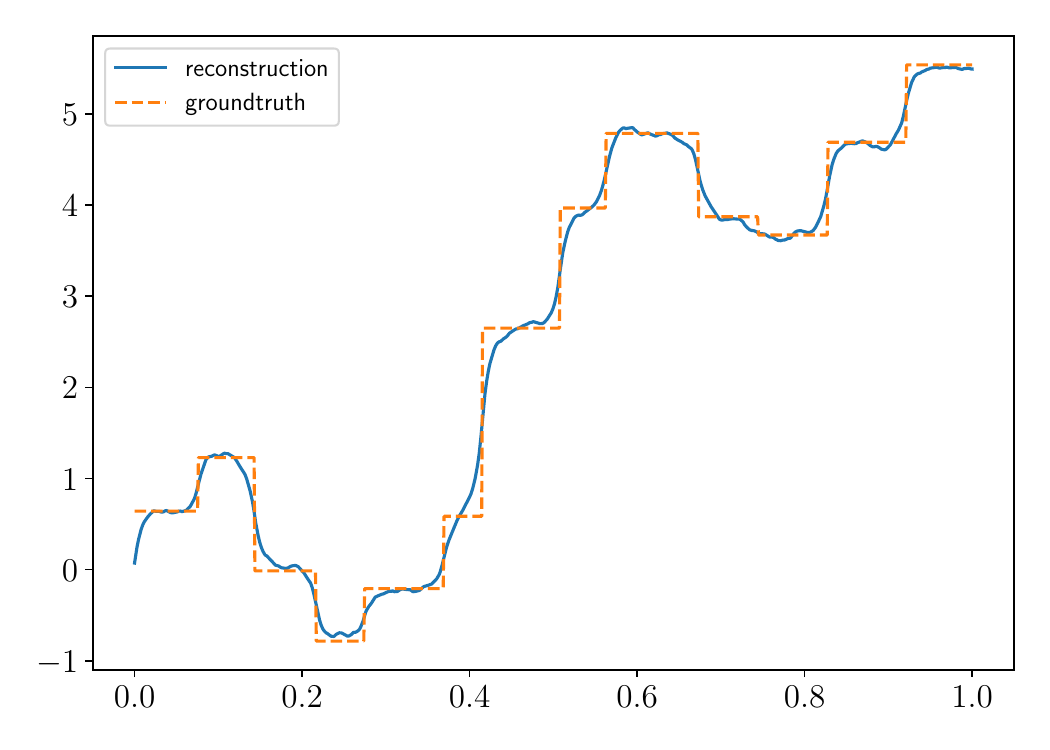}
        \caption{Tikhonov reconstruction with $\alpha = 10^2$}
    \end{subfigure}
    \begin{subfigure}[b]{0.4\textwidth}
        \centering
        \includegraphics[width=1\textwidth]{figures/tik_large.pdf}
        \caption{Tikhonov adversarial reconstruction with $\alpha = 10^{-7}$}
    \end{subfigure}
    \begin{subfigure}[b]{0.4\textwidth}
        \centering
        \includegraphics[width=1\textwidth]{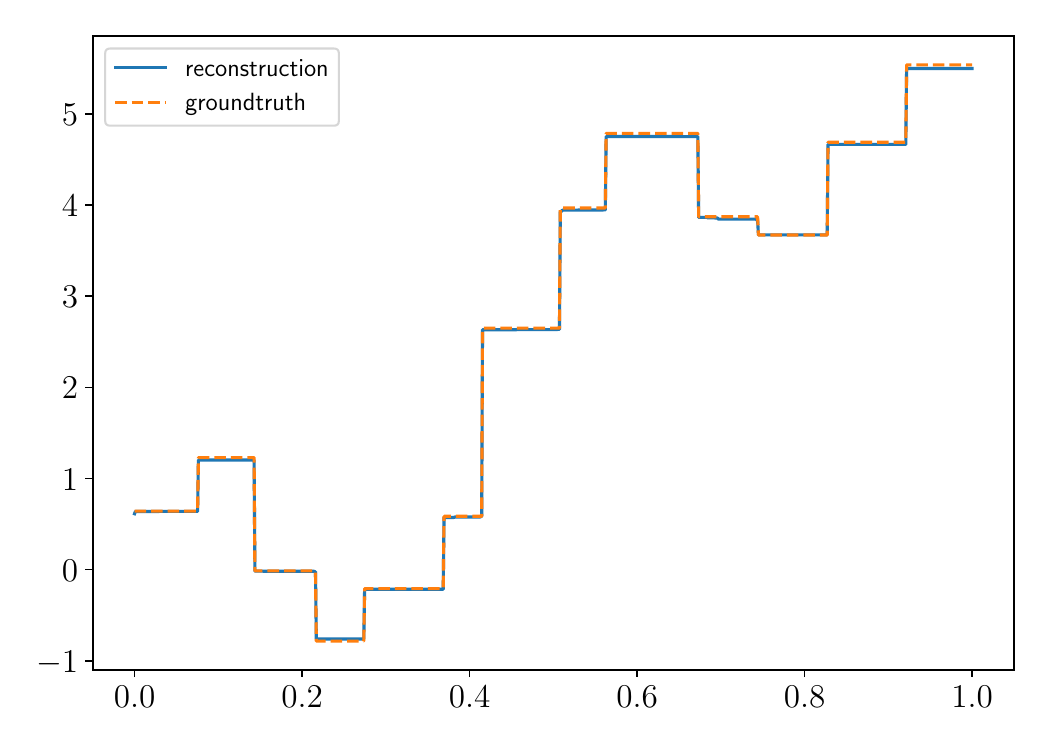}
        \caption{TV reconstruction}
    \end{subfigure}
    \begin{subfigure}[b]{0.4\textwidth}
        \centering
        \includegraphics[width=1\textwidth]{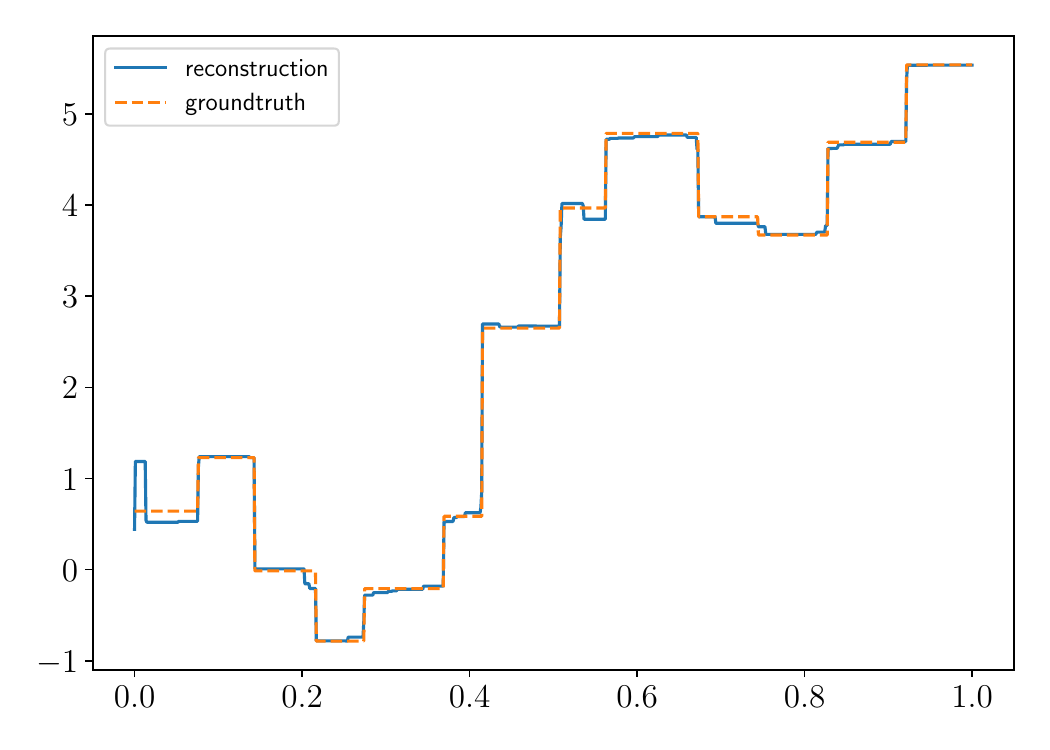}
        \caption{TV adversarial reconstruction}
    \end{subfigure}
    \caption{Results of the total-variation-based reconstruction of 1D-signals in a compressed sensing setting.}
    \label{fig:reconstructions_var}
\end{figure}

As shown in figure \ref{fig:reconstructions_var} (a), this reconstruction, while being able to capture the basic structure of the solution, suffers due to the noisy nature of the measurements, leaving a lot of room for improvement. Additionally, due to the low regularization strength, the approach is very susceptible to adversarial noise, as is clearly visible in figure \ref{fig:reconstructions_var} (b). While $\alpha=10^{-7}$ is a good value for minimizing the $\ell_2$-error in comparison to the ground truth, we also show the results for a much stronger regularization (figure \ref{fig:reconstructions_var} (c) and (d)): While the resulting reconstruction exhibits a clear oversmoothing, it is significantly less affected by the adversarial attack, showing a tradeoff between robustness and fidelity. 

We continue by considering the regularization via total variation \cite{rudin1992nonlinear}, 
\begin{equation}
\label{eq:tv_min}
    \hat{u}_{TV} = \arg\min_u \frac{1}{2}\|Au-f\|_2^2 + \alpha \text{TV}(u),
\end{equation}
where $\text{TV}(u)$ represents the total variation of $u$, and $\alpha$ is a regularization parameter. We employ the alternating directions method of multipliers (ADMM) for solving \eqref{eq:tv_min}. Figure \ref{fig:reconstructions_var} (e) shows that this approach can reconstruct the ground truth nearly perfectly, exhibiting only minor deviations, possibly due to its bias or the mean-seeking behavior of the total variation regularization. While the reconstruction after an adversarial attack (Fig.\ref{fig:reconstructions_var} (f)), is not altered too severely, there are some clear deviations visible. While these deviations can cause a noticeable change in the $\ell_2$-norm in comparison to the reconstruction without adversarial attack, one can see that all jumps (and jump directions) of the ground truth solution are preserved, which is what one would expect from a small symmetric Bregman distance with respect to the total variation, c.f. \eqref{eq:stability_tv1}.

\begin{figure}
    \centering
    \includegraphics[width=0.5\textwidth]{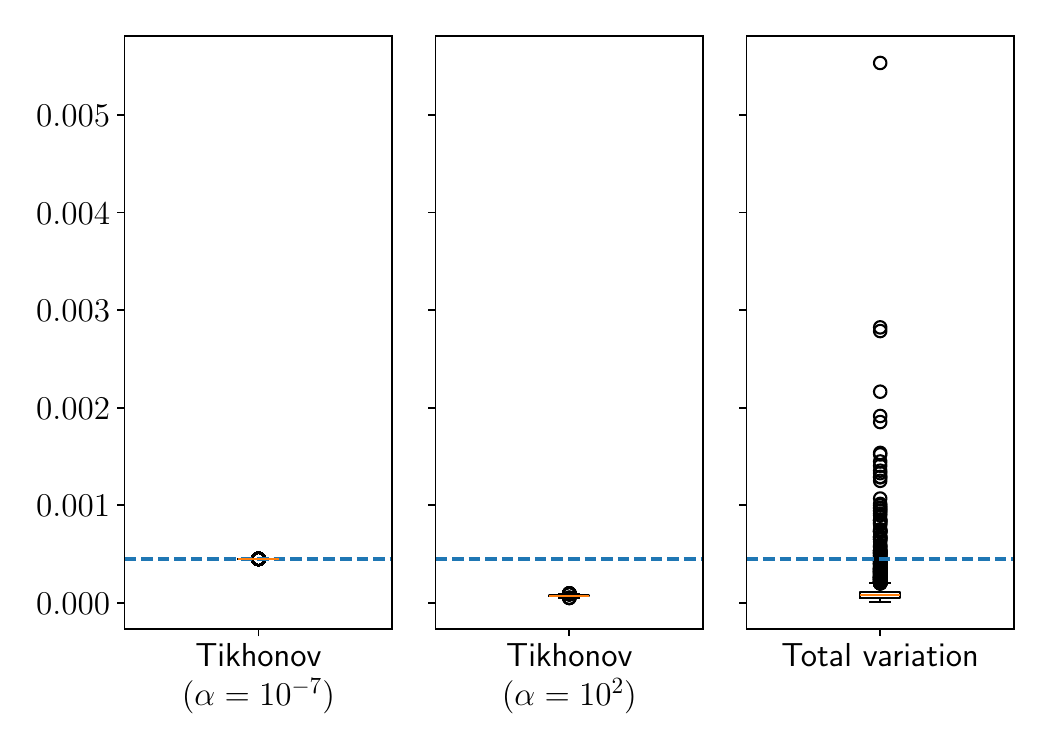}
    \caption{Results of the empirical evaluation of the bound in equation \ref{eq:stability_tv} for different variational reconstruction methods using Gaussian noise. Violations are artifacts of limited precision calculations.}
    \label{fig:bounds_gauss}
\end{figure}

\begin{figure}
    \centering
    \includegraphics[width=0.35\textwidth]{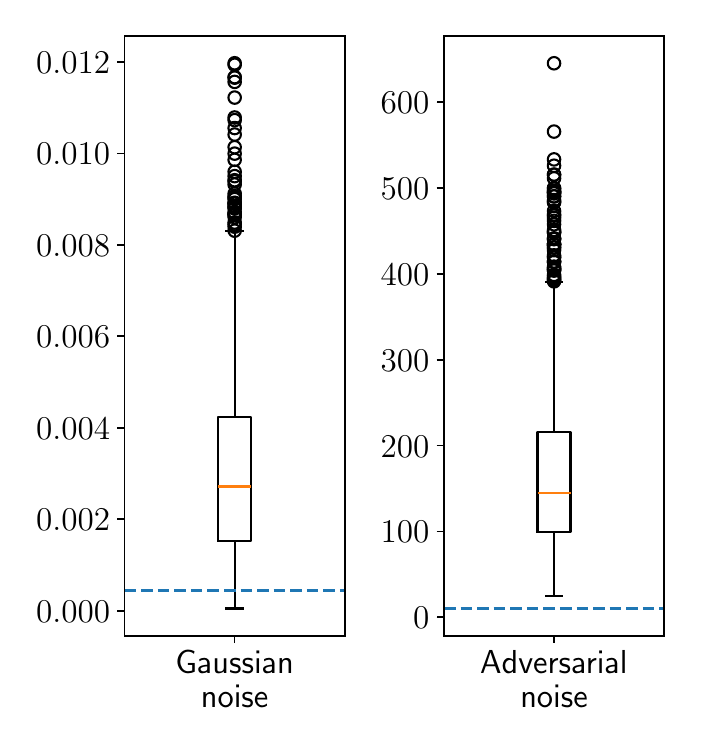}
    \caption{Results of the empirical evaluation of the bound in equation \ref{eq:stability_tv} for Tikhonov regularized reconstruction ($\alpha = 10^{2}$) applied to total variation reconstruction, exhibiting obvious violations.}
    \label{fig:bounds_tv_tik}
\end{figure}

\begin{figure}
    \centering
    \includegraphics[width=0.5\textwidth]{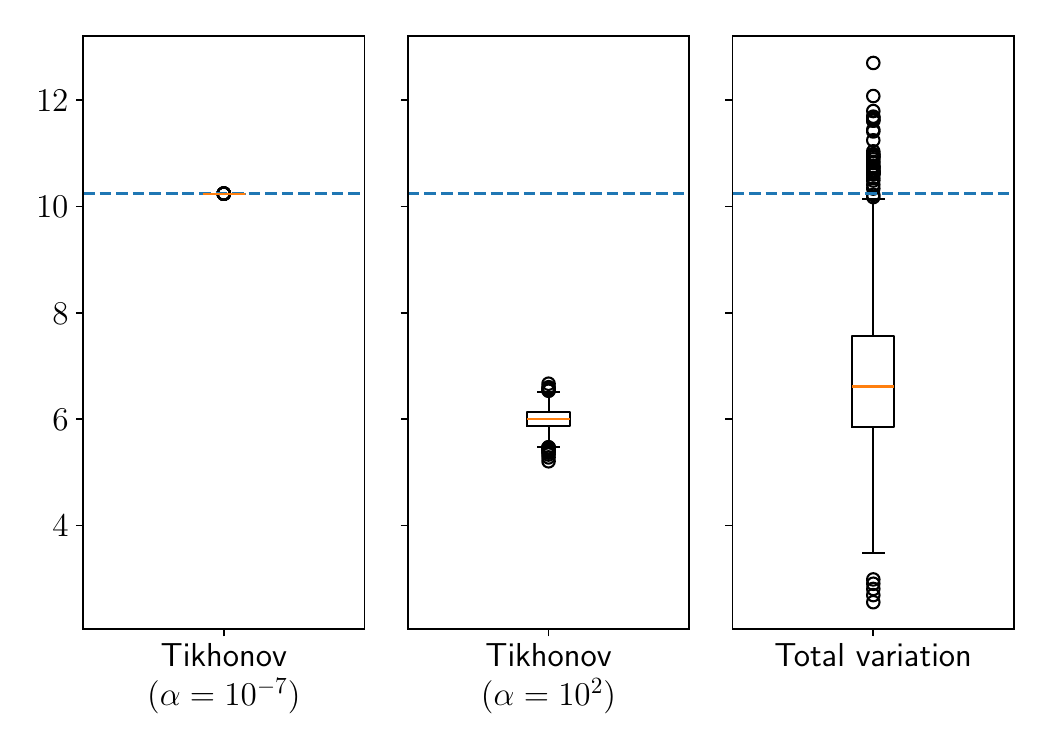}
    \caption{Results of the empirical evaluation of the bound in equation \ref{eq:stability_tv} for different variational reconstruction methods using adversarial noise. Violations are artifacts of limited precision calculations.}
    \label{fig:bounds_adv}
\end{figure}

We further verify empirically that the bound in equation \ref{eq:stability_tv} holds by calculating it over a test dataset, and show the results in figures \ref{fig:bounds_gauss} and \ref{fig:bounds_adv}, for both adversarial and white Gaussian noise and Tikhonov and total variation regularized reconstructions. In all cases, the calculated deviation lies close to or below the bound (shown here as a dashed line). The bound is dictated by the size $\epsilon$ of the projection step used during the adversarial attack, describing the maximal distance between the original and the adversarial sample. We emphasize that the results, while seemingly violating the bounds, all lie close to or below the bound, while the remaining deviations can be attributed to the limits of the numerical precision available. The dependence of equation \ref{eq:stability_tv} on the subgradient of the regularizer leads to different notions of stability for different reconstruction approaches. We show a concrete example in figure \ref{fig:bounds_tv_tik}, visualizing that a total variation reconstruction tends to violate the bounds for a Tikhonov reconstruction.

While the bound, especially the Bregman distance, is not applicable for neural networks trained for image recovery, measuring robustness in terms of measurement consistency $\left\|A\net_{\theta}(f+\delta_{\text{adv}})-f\right\|_{2}$ is interesting as there can be multiple solutions to ill-posed problems which satisfy a similar level of consistency, even when there is a large discrepancy between them in terms of $\ell_2$ error in image space.
 \cite{gandikota2023evaluating}  find that adversarial reconstructions are remarkably stable in terms of measurement consistency, even when there is a significant degradation in the quality of reconstructions. This, again, hints at the fact that adversarial attacks happen in subspaces corresponding to small singular values of the forward operator, such that attacks can exploit any under-regularization. The work further demonstrates universal attacks are feasible and also transferable across different recovery networks showing the potential of black box attacks on image recovery. 
 
\begin{figure}
    \centering
    \begin{subfigure}[b]{0.4\textwidth}
        \centering
        \includegraphics[width=1\textwidth]{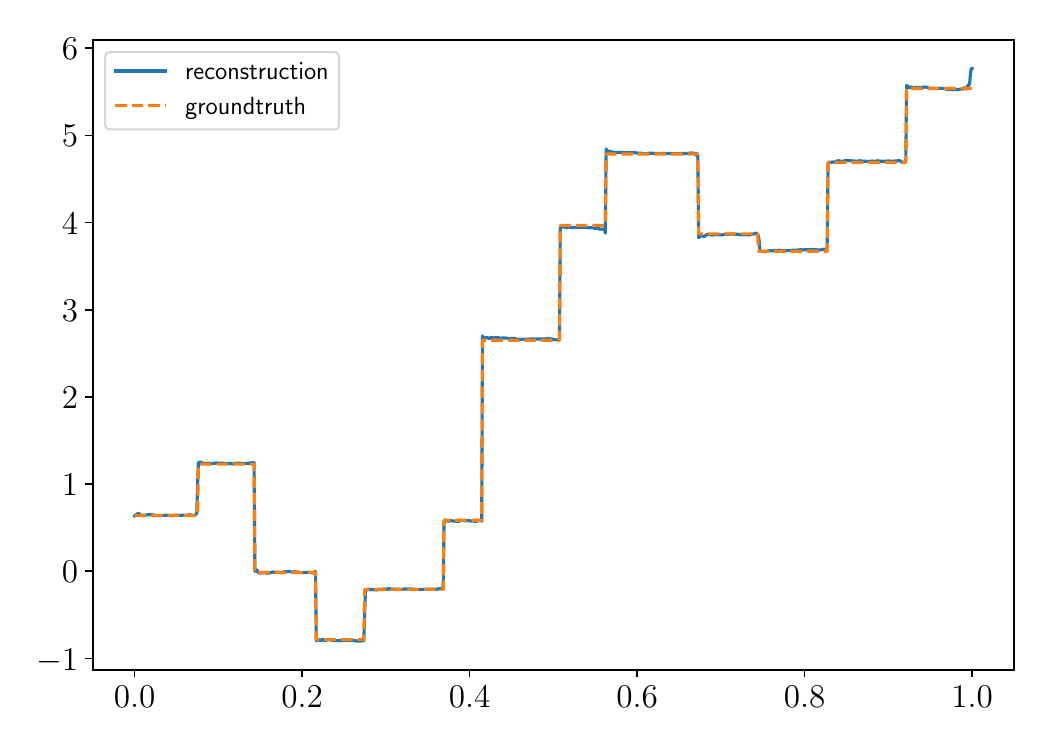}
        \caption{U-Net reconstruction}
    \end{subfigure}
    \begin{subfigure}[b]{0.4\textwidth}
        \centering
        \includegraphics[width=1\textwidth]{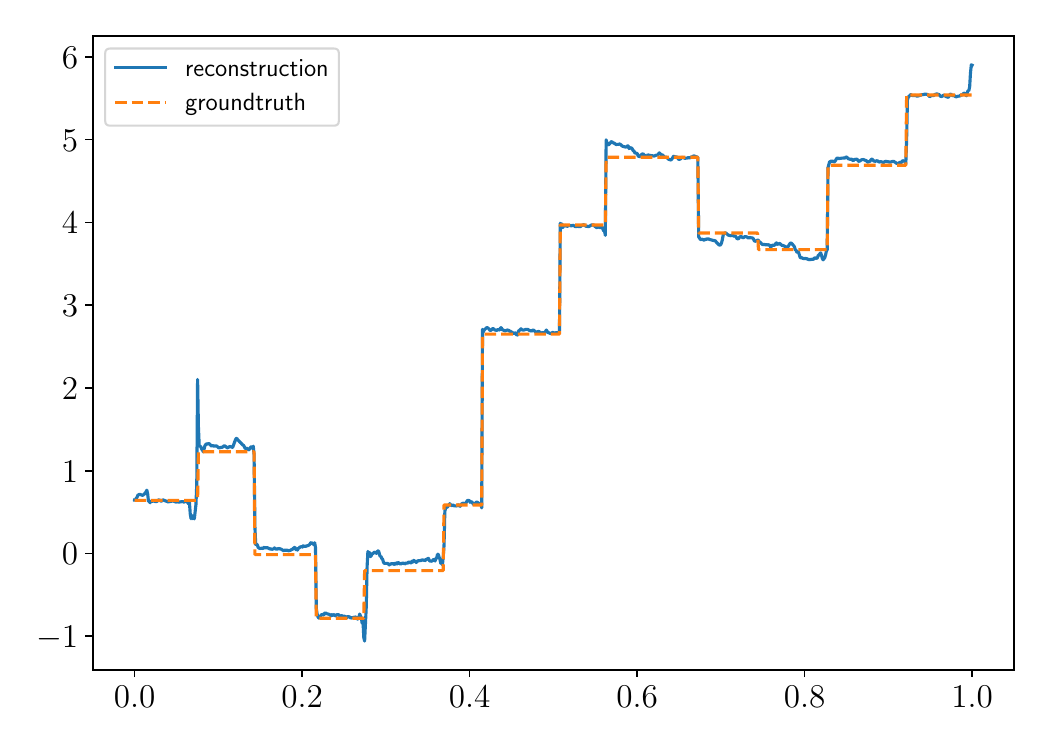}
        \caption{U-Net adversarial reconstruction}
    \end{subfigure}
    \begin{subfigure}[b]{0.4\textwidth}
        \centering
        \includegraphics[width=1\textwidth]{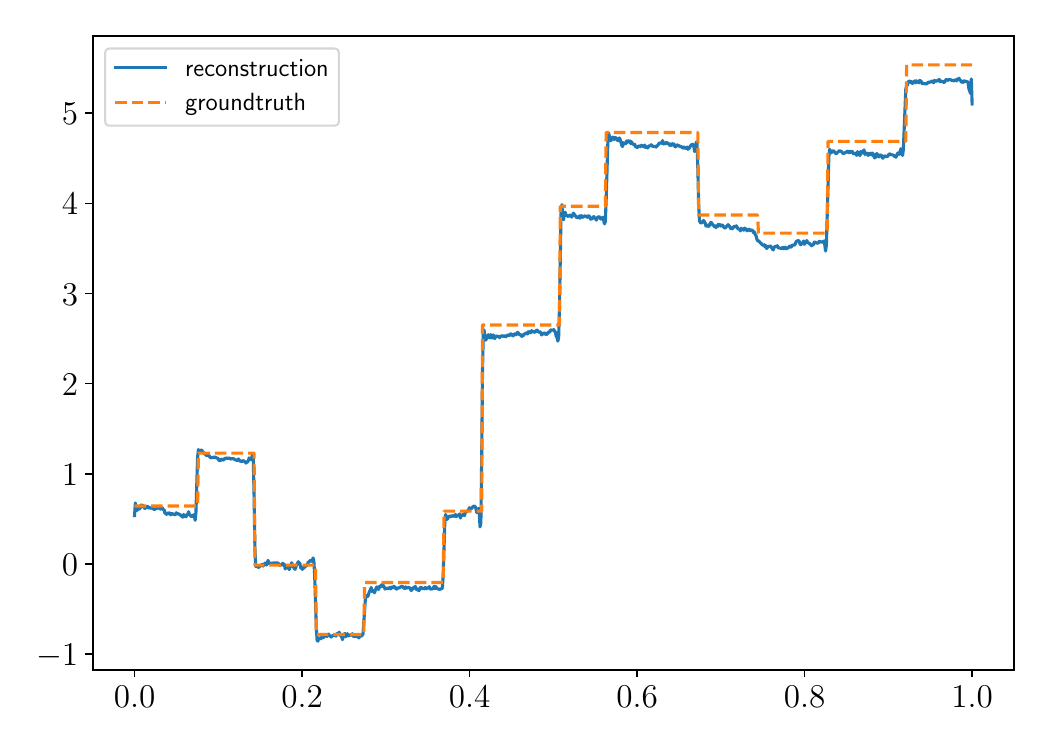}
        \caption{Tiramisu reconstruction}
    \end{subfigure}
    \begin{subfigure}[b]{0.4\textwidth}
        \centering
        \includegraphics[width=1\textwidth]{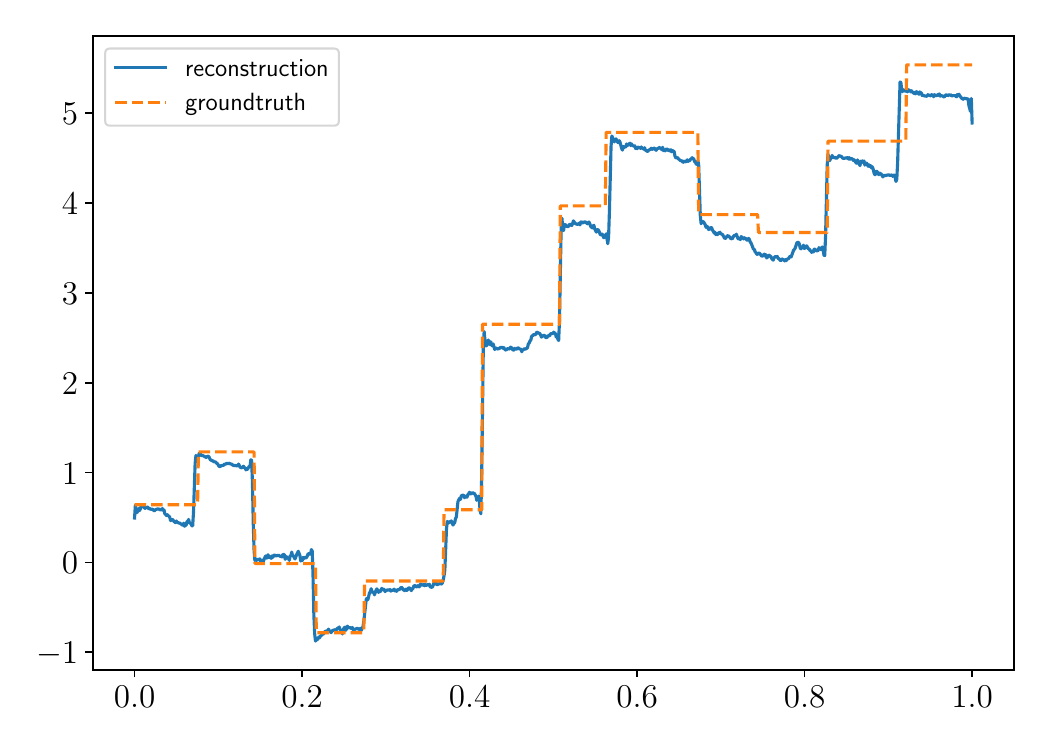}
        \caption{Tiramisu adversarial reconstruction}
    \end{subfigure}
    \begin{subfigure}[b]{0.4\textwidth}
        \centering
        \includegraphics[width=1\textwidth]{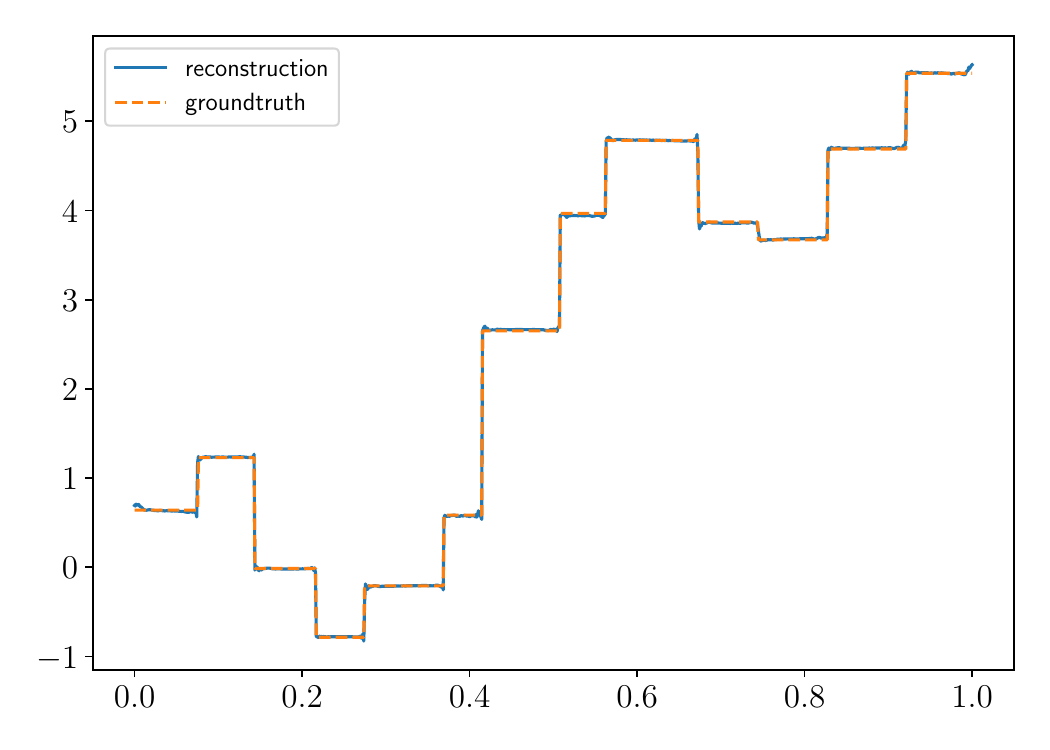}
        \caption{ItNet reconstruction}
    \end{subfigure}
    \begin{subfigure}[b]{0.4\textwidth}
        \centering
        \includegraphics[width=1\textwidth]{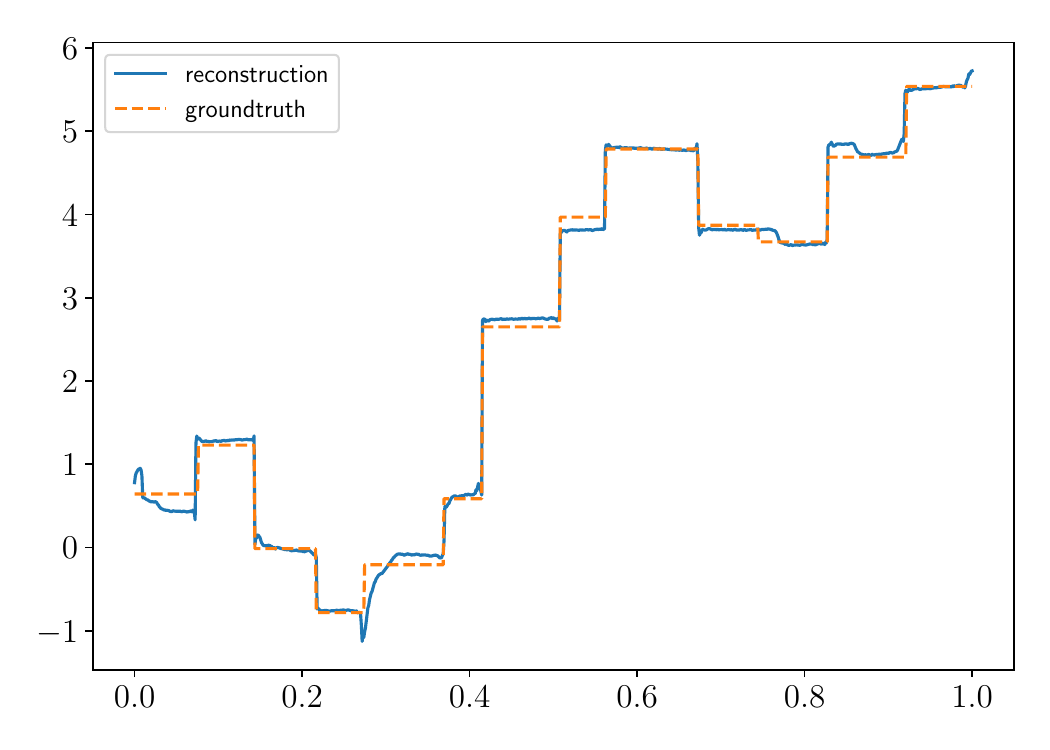}
        \caption{ItNet adversarial reconstruction}
    \end{subfigure}
    \caption{Results of the learned reconstructions of 1D-signals in a compressed sensing setting.}
    \label{fig:reconstructions_networks}
\end{figure}

For completeness, we also show the performance of the neural network-based methods from Genzel \etal \cite{genzel2020solving} in terms of the aforementioned measurement consistency. All learned approaches are trained on a dataset of 8192 samples with additive white gaussian noise (AWGN) with mean $0$ and a standard deviation of $0.03$ ($\mathbb{E}(\|n\|^2) = 0.6785$, with $n$ denoting the noise).
First, we show the reconstruction capability of a U-Net architecture, where the U-Net serves as a post-processor, refining the image obtained from an initial Tikhonov regularized solution, transformation,
\begin{equation}
    \hat{u} = \mathcal{G}_{\hat{\theta}}(\hat{u}_{Tik}),
\end{equation}
where $\mathcal{G}_\theta$ represents a U-Net model.  In Figure \ref{fig:reconstructions_networks} (a) and (b) the resulting reconstruction as well as the reconstruction of an adversarial example are shown. The U-Net reconstruction is comparable in quality to the total variation, while the adversarial example (obtained using the same attack and hyperparameters as in the total variation case), has a somewhat stronger influence on the network performance.

In the next approach, the reconstruction is predicted by a Tiramisu model, where the network is responsible for improving the results of a learned linear forward operator:
\begin{equation}
    \hat{u} = \mathcal{G}_{\hat{\theta}_1}(L_{\hat{\theta}_2}f).
\end{equation}
Here $\mathcal{G}_\theta$ denotes a neural network based on the Tiramisu architecture \cite{jegou2017one}, while $L_\theta$ represents a learned linear transformation which is intended to substitute the Tikhonov reconstruction operator used in the U-Net approach. The achieved reconstruction quality (figure \ref{fig:reconstructions_networks} (c) and (d)) in the normal and the attacked case, after comparable training efforts, is visibly worse than in any other approach. This demonstrates the influence of model information on the reconstruction process as well as the difficulty of training modern neural network architectures, due to the amount of resources required for training and fine-tuning hyperparameters.

Lastly, we reconstruct the signal using a Plug-and-play approach (ItNet), where the U-Net architecture replaces the proximal step in a proximal gradient descent approach 
\begin{equation}
    \hat{u} = u^I, \quad u^{k+1} =  \mathcal{G}_{\hat{\theta}}\left(u^k - \tau A^T\left(Au^k-f\right)\right), \quad u^0 = u_{Tik}
\end{equation}
where $\mathcal{G}_\theta$ again represents a U-Net model, applied as $\operatorname{prox}$-operator in $I$ many proximal gradient descent steps. The reconstruction quality achieved is similar to that of the post-processing U-Net (see figure \ref{fig:reconstructions_networks} (e) and (f)). The attacked reconstruction exhibits artifacts comparable to a weak form of the artifacts in the U-Net case, demonstrating in another way the effects of the amount of incorporated model information.

\begin{center}
    \begin{table}[t]
    \centering
    \caption{Quantitative results of the different reconstruction methods. In the rows below, $\hat{u}$ denotes the reconstruction on noisy data $f$ (which is the same for all methods). $\hat{u}_{adv}$ denotes the reconstruction on (method-specific) data $f_{adv}$ computed via an adversarial attack on each method by trying to maximize $\|\hat{u}_{adv} - u_{gt}\|$ with one FGSM step. We report the reconstruction quality and data fidelity for both, the normal and adversarially attacked solutions, as well as the difference between the two solutions in an $\ell_2$ sense, in the symmetric Bregman distances of the Tikhonov regularization, and in terms of their difference in data space, i.e., after applying the forward operator.
    }
    \begin{tabular*}{500pt}{@{\extracolsep\fill}lcccccc@{\extracolsep\fill}}
        \toprule
        & \textbf{Tikhonov ($\alpha = 10^{-7}$)}& \textbf{Tikhonov ($\alpha = 10^{2}$)} & \textbf{TV} & \textbf{U-Net} & \textbf{Tiramisu} & \textbf{ItNet} \\
        \midrule
        $\|\hat{u}-u_{gt}\|^2$          &  4.39      & 30.85          &  0.66          &  0.38 &  8.68 & \textbf{0.34} \\
        $\|\hat{u}_{adv}-u_{gt}\|^2$    & 86.37      & 51.45          & \textbf{11.54} & 25.51 & 46.97 & 23.57 \\
        
        $\|A\hat{u}-f\|^2$              & \textbf{0} & 31.44          &  0.84          &  0.46 & 10.68 &  0.38 \\
        $\|A\hat{u}_{adv}-f_{adv}\|^2$  & \textbf{0} & 27.73          &  7.67          & 26.29 & 41.42 & 24.9 \\
        
        $\|A\hat{u}-A\hat{u}_{adv}\|^2$ & 20.48      & \textbf{6.91}  & 10.34          & 21.43 & 28.45 & 20.17 \\
        $\|D\hat{u}-D\hat{u}_{adv}\|^2$ & 30.41      &  0.03          &  1.62          &  3.85 &  5.47 &  4.9 \\
        
        $\|\hat{u}-\hat{u}_{adv}\|^2$   & 66.74      & \textbf{6.52}  &  9.28          & 22.26 & 25.75 & 20.74 \\
        $\|f-f_{adv}\|^2$               & 20.48      & 20.48          & 20.48          & 20.48 & 20.48 & 20.48 \\
        \bottomrule
    \end{tabular*}
    \label{fig:recon_comp_quant}
    \end{table}
\end{center}

Table \ref{fig:recon_comp_quant} shows the quantitative results of our evaluation in a variety of relevant metrics for each approach with noisy measurements as well as adversarial examples.  We consider the following metrics: reconstruction quality in terms of proximity to ground truth for both noisy inputs $\|\hat{u}-u_{gt}\|^2$, and adversarial examples $\|\hat{u}_{adv}-u_{gt}\|^2$, measurement consistency of the reconstructions for noisy inputs $\|A\hat{u}-f\|^2$ and adversarial examples $\|A\hat{u}_{adv}-f_{adv}\|^2$. Obtaining both good reconstruction quality and measurement consistency are crucial for reconstruction algorithm. In addition to these metrics, we also evaluate $\|D\hat{u}-D\hat{u}_{adv}\|^2$  for characterizing the theoretical bound in the Tikhonov case, which also empirically shows the trade-off between robustness and reconstruction quality. Finally, we measure the deviation between the noisy measurements and adversarial inputs $\|f-f_{adv}\|^2$, and corresponding deviations in the reconstructions $\|\hat{u}-\hat{u}_{adv}\|^2$.
The ratio of these measures $\|\hat{u}-\hat{u}_{adv}\|^2$ and $\|f-f_{adv}\|^2$ describes the Lipschitz-constant of reconstruction algorithm.  

The best performance in each of the metrics described is marked in bold in  Table \ref{fig:recon_comp_quant}.
In terms of reconstruction quality with noisy inputs, we observe the best results with ItNet, followed by the postprocessing U-Net approach and the TV reconstruction. Tikhonov regularized reconstructions are acceptable for good choices of $\alpha$, while large values of $\alpha$ lead to over-regularization and bad reconstruction quality.  This highlights the superiority of learned approaches regarding reconstruction quality, purely in terms of the "usability" of the end-result, with no regard to (hyper-)parameter count or compute requirements. 

To summarize: In our experiments, we analyzed common variational, model-based approaches such as TV and Tikhonov regularized reconstruction on the one hand, and on the other hand, learned approaches incorporating varying degrees of model information (following the experiments in \cite{genzel2020solving}).
We also found a tradeoff between reconstruction quality and the robustness of the approach to adversarial attacks. Additionally, we showed that it is possible to empirically verify the bounds given by equation \ref{eq:stability_tv}. We would like to note, however, that this bound is regularizer-dependent, resulting in varying notions of robustness. These robustness measures are not consistent (e.g. TV solutions might violate the robustness measure of the Tikhonov regularization). Learned approaches are also harder to quantify in terms of bounds, leading to again further robustness measures.

\paragraph{Targeted changes} In addition to untargeted attacks which aim to degrade the quality of reconstructions, it is also interesting to evaluate the robustness to targeted attacks which trigger the reconstruction method to produce a specific, (realistic) reconstruction.  \cite{pmlr-v121-cheng20a} perform adversarial attacks to generate tiny features, which cannot be recovered well by MRI reconstruction networks, and propose adversarial training to improve the network's sensitivity to such features.   \cite{darestani2021measuring, morshuis2022adversarial,huang2018some} show that adversarial perturbations can alter diagnostically relevant regions.
In \cite{gandikota2023evaluating} the authors demonstrate that localized adversarial attacks targeting diagnostically relevant regions can recover diagnostically different images even with extremely small perturbations such that resulting solutions still maintain a high degree of measurement consistency. While this appears to be contradictory to \cite{gandikota2022adversarial}, where drastic changes that are highly inconsistent with the measurements were obtained after small adversarial changes of the input data, differences in the knowledge of the forward operator could be an explanation for the vastly different behaviors: While \cite{gandikota2023evaluating} considered the same forward operator for all instance of the data set, \cite{gandikota2022adversarial} considered variable and even unknown forward operators.

\paragraph{Defense}
The simplest defense to deal with additive perturbations is training with additive noise.
\cite{krainovic2023learning,genzel2020solving} show that training with noise improves the adversarial robustness of reconstruction networks, with \cite{krainovic2023learning} showing this to be the optimal strategy for training robust denoisers.
Prior works \cite{pmlr-v119-raj20a,agnihotri2023unreasonable,Choi_2020_ACCV,castillo2021generalized} also perform adversarial training \cite{aleksandermadry} or regularization to improve robustness.
While adversarial training can improve robustness when the attack is (roughly) known, yet even this does not "guarantee" robustness.
Further, improved robustness through adversarial training leads to reduced quality reconstruction. On the other hand, high reconstruction quality invariably comes at a cost of reduced robustness, we refer to  \cite{ohayon2023perception,gottschling2020troublesome} for a discussion on this trade-off.

\subsection{Robustness to distribution shifts}
The work \cite{darestani2021measuring} studies the effect of distribution shifts due to different acquisition techniques, different anatomies, and 
difficult-to-reconstruct samples as evaluated by a state-of-the-art method. The authors find that the performance drop on distribution shifts is similar for
 trained and untrained methods (e.g. model-based approaches or untrained neural network priors).
 Untrained methods using hyperparameters tuned for a particular distribution do not perform as well with distribution shifts. Further theoretical analysis is performed in \cite{9940488}, where explicit error bounds for mismatched CNN-priors for steepest descent RED are derived. \cite{pmlr-v162-darestani22a}  propose to fix the effects of distribution shifts through a self-supervised domain adaptation method paired with inference-time training to improve the robustness to distribution shifts.
 
 \subsection{Robustness to changes in forward measurement operator}

 Another desirable property of an image reconstruction algorithm is the robustness to changes in the measurement model. Classical variational approaches allow modifications, for example, changes in the noise model or modifications of the forward operator $A$, as they can easily be incorporated into the energy minimization by appropriate changes of the energy function. While this also holds for learned regularizers, denoising priors, or generative priors, end-to-end trained neural networks, including the model-based unrolled networks suffer from a lack of adaptivity. This
means that a network trained for a specific forward operator $A$ and noise model suffers from a significant performance drop if these are modified, and therefore have to be retrained for the new measurement model  \cite{antun2019instabilities}. To address this, \cite{gilton2021model} propose a fine-tuning-based as well as a training-free approach to adapt trained models to variations in forward operator, whereas \cite{gossard2022training} propose training with different forward operators. \cite{deq_model_robust} show that unrolled networks based on deep equilibrium models \cite{gilton2021deep} are robust to changes in the measurement model. A few methods 
\cite{nan2020deep,wang2022modeling,9546648}  account for uncertainty in the forward operator explicitly in the network to improve robustness to errors in calibrating the forward measurement process.

\subsection{Robustness in recovering fine details} The authors of \cite{antun2019instabilities} find that different trained networks have different degrees of robustness in recovering fine details not seen in training data, ranging from the complete removal of such details to their faithful recovery.
\cite{darestani2021measuring} observe that this ability to recover fine details is directly correlated with the overall reconstruction performance, and improving it also improves the ability to recover fine details. 
\cite{antun2019instabilities} consider fine details not belonging to the null space of the forward operator. As network hallucinations, changes and removal of details are common problems encountered in learning-based approaches, research focussing on the enforcement of data-consistent solutions has emerged, e.g. \cite{moeller2019controlling}, where a gradient descent algorithm utilizing network-predicted descent directions is modified to converge globally to the minimizer of the data fidelity. Yet, when certain details belong to the null space of the forward operator, the problem of recovering them is rather a generative task and leads to the desire to be able to draw realistic possible sample reconstructions or to actively \textit{exploit} solutions with certain properties. We will briefly summarize the former before providing some more details on the latter. 

\subsection{Robustness for Bayesian Methods}
Instead of the recovery of a single solution and the investigation of how the single predicted solution changes as the measurements change, the perspective of \textit{Baysian inverse problems} is that the prediction of the posterior should exist, be unique, and be locally Lipschitz continuous for changing data (c.f. \cite{doi:10.1137/23M1556435}). Consequently, the terms continuity and stability depend on a suitable choice of distance between probability measures and can yield well-posed problems far more often than in the variational setting (see \cite{doi:10.1137/23M1556435}). In finite dimensions, this effect can be understood by relating energy minimization methods to the Bayesian setting via maximum a-posterior probability estimates. Naturally, one has to expect that $\arg\min_u -\log p(u|f)$ can be discontinuous even if $p(u|f)$ depends on $f$ continuously, and we refer to \cite{altekruger2023conditional} for a nice example. We also refer to \cite{altekruger2023conditional} for proving that the Lipschitz continuity of the conditional generative model transfers to a stability estimate for the posterior, and to the references therein (e.g. \cite{gouk2021,miyato2018spectral,Hagemann_2021,salmona2022can}) for further discussions on the trade-off between the regularity and the expressivity of (conditional) generators.

\section{Explorability}
In the case where many of the singular values of the forward operator are either zero or very small in comparison to the expected noise level of the measurements, any reconstruction method has to select a solution from many possible choices. For instance, MAP estimates try to select the most probable one, and Bayesian methods allow picking multiple ones by sampling from the estimated posterior. Yet, considering the high dimensionality of the underlying space as well as the risk of complex (not well-localized) posteriors, a very large number of samples could be necessary to get a good impression of the variety of possible reconstructions. Among such samples, many will be similar, and - depending on the application - only a few of them might be relevant to answer an underlying question of interest.
Therefore, some researchers have started focussing on the explorability of inverse reconstruction problems: To provide more control during the reconstruction process, a guiding mechanism can provide solutions that are not only data consistent but also fulfill additional criteria, such as specific semantic interpretations or particular texture properties.

 For instance, \cite{cohen2024from} attempts to address the lack of diversity in posterior sampling using diffusion models, and proposes a guidance mechanism to reduce similarity between outputs starting from different random noises.
 Bahat \etal~\cite{bahat2020explorable} address a limitation in existing super-resolution methods, which typically produce a single high-resolution output from a low-resolution input. The authors propose the composition of a Generative Adversarial Network (GAN) with a function that provably enforces the consistency with the measurements.  
To control the reconstruction,  they introduce a control signal $z$, which is induced into each layer of the neural network and is designed to manipulate image gradients, thereby enabling texture modification.
In \cite{bahat2021s} this work is extended to JPEG image decompression including an option to generate the control $z$ via the optimization over an image classifier to guide the reconstruction towards a specific classification.  
    
Following the approach of learning a classifier to guide the reconstruction, 
Droege \etal~\cite{droege2022bmvc} proposed to explore the space of possible computed tomography reconstruction via an energy minimization technique considering 
\begin{equation}
  \min_{u \in [0,1]^N}   \|A u-f\|^2_2 + \lambda H(C_{\theta}(u)-d)
\label{eq:explorable_droege}    
\end{equation}
with $H$ denoting a suitable loss function, $C_{\theta}$ denoting a robust (fixed) classifier, and $d$ being a guidance class parameter. As an application, the reconstruction of computerized tomography images of the human lung with different levels $d$ of predicted malignancy of (localized) nodules is presented. 

\begin{figure}
    \centering
    \begin{subfigure}[b]{0.19\textwidth}
        \centering
        \includegraphics[width=1\textwidth]{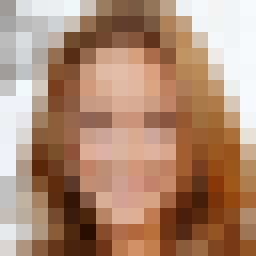}
        \caption{Low-res}
    \end{subfigure}
    \begin{subfigure}[b]{0.19\textwidth}
        \centering
        \includegraphics[width=1\textwidth]{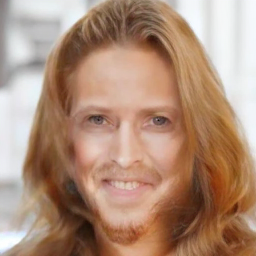}
        \caption{}
    \end{subfigure}
    \begin{subfigure}[b]{0.19\textwidth}
        \centering
        \includegraphics[width=1\textwidth]{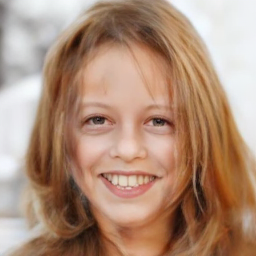}
        \caption{}
    \end{subfigure}
    \begin{subfigure}[b]{0.19\textwidth}
        \centering
        \includegraphics[width=1\textwidth]{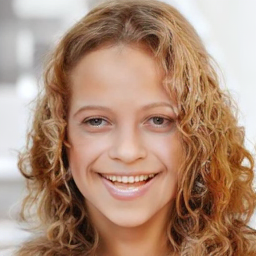}
        \caption{}
    \end{subfigure}
    \begin{subfigure}[b]{0.19\textwidth}
        \centering
        \includegraphics[width=1\textwidth]{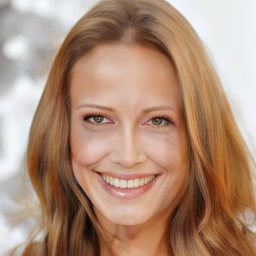}
        \caption{}
    \end{subfigure}
    \caption{Exploring solutions to $16\times$ super-resolution through text using method from \cite{gandikota2023exploring}. The text prompts used are ''a high resolution photograph of a face of b) a man c) a child d) a smiling child with curly hair e) a smiling woman''.}
    \label{fig:text_explore}
\end{figure}

Recently, Gandikota and Chandramouli \cite{gandikota2023exploring}  introduced zero-shot text-guided exploration of solutions to super-resolution using text-to-image diffusion models \cite{ramesh2022hierarchical,saharia2022photorealistic} by adapting different diffusion prior based reconstruction methods \cite{chung2022diffusion,wang2022zero,song2023pseudoinverseguided}. Among these methods, adapting \cite{wang2022zero} resulted in solutions that have high degree of data consistency. This involves ensuring analytical data consistency through projection at every step in the reverse diffusion process following \cite{wang2022zero}:
\begin{equation}
    \hat{u}_{0|t}:=A^{\dagger}f + (I - A^{\dagger}A){u}_{0|t}.
    \label{eq:ndm core}
\end{equation}
where ${u}_{0|t}$ is the MMSE estimate of the clean image at step $t$ of the reverse diffusion process. \cite{gandikota2023exploring}  adapt this to the cascaded diffusion process at different resolutions in \cite{ramesh2022hierarchical} by appropriately modifying the forward operator at each resolution. Figure~\ref{fig:text_explore} exemplifies a result of this approach for the task of $16\times$ super-resolution. 

\section{Conclusions}
We have provided an overview of model-based, learning-based, and hybrid techniques for linear inverse problems with a focus on the robustness of point-based predictors. Our goal was to show that - although the notion of $\ell_2$ stability is dominant in the machine learning literature - at least convex variational methods give rise to provable stability but in a different metric, i.e., the sum of consistency in the data space (after applying the forward operator to the reconstruction) and the symmetric Bregman distance with respect to the used regularizer. To what extent different neural network architectures and training schemes could also lead to different notions of stability, remains an interesting direction of future research. Furthermore, a clear bias-variance (or expressiveness-robustness) trade-off seems to persist. Beyond point-based estimates of solutions, the entire posterior might be difficult to sample from, such that we advertised research in the active (application-specific) exploration of different meaningful and realistic solutions. The latter can include a control for specific classification problems in medical as well as different forms of guidance, including text, for the reconstruction of natural RGB images in challenging situations, with diffusion models being a promising recent technique for representing strong generative priors. 

\section{Data Availability Statement}
The data that support the findings of this study are openly available on Github at \url{https://github.com/AlexanderAuras/GAMM-Overview-23/}.

\section{Acknowledgements}
We acknowledge the support of the German Research Foundation, Project MO 2962/7-1, and Research Unit 5336, "Learning to Sense".

\section{Conflict of interest}
The authors declare no conflict of interest.

\bibliography{bibliography}
\end{document}